\documentclass[sort,compress,11pt]{elsarticle}

\usepackage{amssymb}
\usepackage{amsthm}
\usepackage{amsmath, esint}
\usepackage{mathtools}
\usepackage{mathrsfs}
\usepackage[ruled, lined, linesnumbered, longend]{algorithm2e}
\usepackage{array}
\usepackage{booktabs}
\usepackage{multirow}
\usepackage{listings}
\usepackage{tabu}
\usepackage{enumerate}
\usepackage{lineno}
\usepackage{fullpage}
\usepackage{xcolor}
\usepackage[colorinlistoftodos]{todonotes}
\usepackage[colorlinks=true]{hyperref}
\usepackage{url}
\usepackage{textcomp}

\usepackage{graphicx}
\usepackage{subfig}

\usepackage{makecell}
\usepackage[export]{adjustbox}
\usepackage{caption}
\usetikzlibrary{shapes}
\biboptions{numbers,comma,round,square}
\usepackage[para,online,flushleft]{threeparttable}

\usepackage{orcidlink}

\definecolor{myblue}{RGB}{63, 144, 218}
\definecolor{myorange}{RGB}{221, 132, 82}
\definecolor{mygreen}{RGB}{85, 168, 104}
\definecolor{myyellow}{RGB}{255, 169, 14}
\definecolor{myred}{RGB}{189, 31, 1}

\graphicspath{ {./figs/} }

\journal{Elsevier}
\DontPrintSemicolon
\definecolor{orcidlogocol}{HTML}{A6CE39}
\begin{document}
\makeatletter
\def\ps@pprintTitle{%
  \let\@oddhead\@empty
  \let\@evenhead\@empty
  \let\@oddfoot\@empty
  \let\@evenfoot\@oddfoot
}
\makeatother
\begin{frontmatter}

\title{Patched Flow Matching: Generative Wall-Pressure Reconstruction Beyond Training-Domain Scales from Sparse Sensors}

\author[cMAE]{Meet Hemant Parikh\orcidlink{0009-0008-1284-6368}}
\author[cMAE,ndAME]{Yi Liu\orcidlink{0000-0002-9174-2004}}
\author[cMAE,ndAME]{Jian-Xun Wang\orcidlink{0000-0002-9030-1733}}
\address[cMAE]{Sibley School of Mechanical and Aerospace Engineering, Cornell University, Ithaca, NY}
\address[ndAME]{Department of Aerospace and Mechanical Engineering, University of Notre Dame, Notre Dame, IN}

\begin{abstract}
Characterizing the complete wall-pressure spectrum in turbulent wall-bounded flows requires simultaneous access to the viscous-scale high-wavenumber content and the outer-layer low-wavenumber content---a requirement that neither short-domain direct numerical simulation (DNS) nor sparse experimental
measurements alone can satisfy. We propose Patched Flow Matching (Patched FM), a generative framework that fuses these two complementary sources by learning a patch-local prior over inner-scaled wall-pressure statistics from short-domain DNS and assimilating sparse sensor measurements at inference time through training-free posterior sampling. The patch-additive decomposition of the flow matching vector field decouples the generative prior
from the global domain size, enabling reconstruction on domains arbitrarily larger than the training configuration. By expressing the patch prior in inner-scaled coordinates, where high-wavenumber wall-pressure statistics are approximately Reynolds-number invariant, the framework extends to higher Reynolds numbers through hierarchical transfer learning with as few as $500$ short-domain snapshots ($2.5\%$ of the base training data) at a fraction of
the scratch-training cost. Applied to compressible channel-flow DNS at $Re_\tau = 180$, $500$, and $1000$, Patched FM reconstructs full-resolution wall-pressure fields on a domain four times larger than the training configuration ($L_x^L = 16\pi\delta$ versus $L_x^S = 4\pi\delta$) from sensor coverage as low as $0.25\%$, recovering the low-wavenumber spectral content inaccessible to short-domain DNS with high fidelity in both streamwise and spanwise
directions. Zero-shot generalization to unseen Reynolds numbers and ablation studies further confirm the role of inner scaling as a physical prerequisite for data-efficient Reynolds-number transfer.
\end{abstract}

\end{frontmatter}

\section{Introduction}

Turbulent wall-bounded flows underpin a vast range of engineering systems, and the fluctuating pressure they exert on the bounding surface is a primary driver of flow-induced noise, structural vibration, and fatigue~\cite{blake2017mechanics,bull1996wall,devenport2018sound}. Accurate characterization of the wall-pressure field across its full spectral range is a prerequisite for aeroacoustic prediction, structural design, and flow control. The challenge is fundamental: the wall-pressure spectrum is broadband and multi-scale, spanning physical mechanisms separated by orders of magnitude in both length and time. The high-wavenumber, high-frequency tail is set by viscous-scale near-wall dynamics and is largely independent of the outer flow; the low-wavenumber content, by contrast, is imprinted by energetic large-scale and very-large-scale motions in the outer layer, whose characteristic streamwise wavelengths scale with the outer length $\delta$ and can reach $\mathcal{O}(10\delta)$ or beyond~\cite{kim1999very,hutchins2007evidence,marusic2010predictive,mathis2009large}. At high Reynolds numbers, these outer-scale motions become increasingly energetic, and simultaneously resolving their low-wavenumber spectral content alongside the viscous-scale high-wavenumber structures---across a single, long, high-fidelity domain---remains a central challenge in wall-pressure characterization~\cite{liu2024subconvective,liu2025reynolds}.

Two established investigative routes exist, each capturing one end of the spectrum while missing the other. Direct numerical simulation (DNS) resolves viscous-scale near-wall structures with full fidelity, but capturing the low-wavenumber outer-scale content demands a streamwise domain commensurate with the $\mathcal{O}(10\delta)$ energetic wavelengths, combined with sufficiently long integration times to converge the slowly decorrelating large-scale statistics~\cite{lozano2014effect}. At high Reynolds numbers this becomes prohibitive: large-domain channel DNS reaches grids in excess of $10^9$ points~\cite{liu2025reynolds} and must be advanced over many eddy-turnover times before the outer-scale statistics are converged. In practice, simulations therefore adopt reduced streamwise domains~\cite{jimenez1991minimal}, which preserve near-wall accuracy but truncate every wavelength beyond the box length: short-domain DNS faithfully resolves the high-wavenumber tail while missing the energetic low-wavenumber content entirely.

Physical experiment faces the complementary limitation. Arrays of flush-mounted pressure transducers can span streamwise extents far exceeding any feasible DNS domain and thus provide access to the energetic large-scale, low-wavenumber content~\cite{farabee1991spectral,klewicki2008statistical,gravante1998characterization}. However, the achievable inter-sensor spacing is orders of magnitude coarser than the viscous length $\ell^* = \nu/u_\tau$, placing the high-wavenumber viscous-scale structures far below the Nyquist wavenumber of any practical transducer array~\cite{hu2022sensor}. Moreover, measured spectra are convolved with the device transfer function, making deconvolution and multiscale recovery difficult \cite{prigent2020deconvolution}. The two sources are thus precisely complementary: DNS provides the high-wavenumber prior but lacks domain extent, while sparse sensor arrays constrain the large-scale field but carry no information about the small-scale structure.

These observations point to a natural strategy: rather than committing to a single source, fuse the complementary information that each provides---a high-wavenumber prior from tractable short-domain DNS, and large-scale constraints from sparse, wide-aperture sensor measurements. Recovering a full-resolution, full-spectrum wall-pressure field from such complementary sources is a data assimilation and field reconstruction problem, for which machine learning has emerged as a powerful and flexible tool. Super-resolution and reconstruction methods based on convolutional neural networks have demonstrated accurate recovery of turbulent fields
from coarse observations, including super-resolution of wall-bounded turbulence~\cite{fukami2019super,fukami2021machine},
sparse reconstruction from limited measurements~\cite{chaurasia2024reconstruction},and prediction of wall-bounded turbulent structures from wall observations~\cite{guastoni2021convolutional,balasubramanian2021predicting}. However, these approaches are fundamentally deterministic, yielding a single point estimate that cannot capture two distinct sources of non-uniqueness inherent to the reconstruction problem. The first is physical: turbulence is an inherently stochastic process, and a given set of sparse sensor measurements is consistent with an ensemble of physically realizable instantaneous fields that differ in their small-scale structure. The second is mathematical: full-field reconstruction from sparse observations is a severely ill-posed inverse problem, for which multiple fields are equally consistent with the data. A deterministic model collapses both sources of non-uniqueness to a single blurred estimate, suppressing coherent small-scale structures and underestimating extreme events. A probabilistic
generative framework is therefore not merely desirable for uncertainty quantification, but essential for faithfully representing the stochastic character of turbulent wall-pressure fields.

Score-based diffusion models~\cite{song2020score} and flow-matching models~\cite{lipman2022flow,albergo2025stochastic} learn expressive priors over high-dimensional fields and naturally produce ensembles of physically consistent samples. Posterior-sampling algorithms~\cite{chung2022diffusion,song2023pseudoinverse,parikh2026d} further enable training-free conditioning on arbitrary observation operators at inference time, decoupling the learned prior from any specific measurement configuration---a critical property when sensor layouts vary across experimental
setups. These capabilities have been exploited across a growing range of turbulence applications: diffusion and flow-matching models have been applied to super-resolution and synthesis of turbulent velocity fields~\cite{gao2024bayesian,du2024conditional,fan2025neural,chakraborty2026adaptive,vishwasrao2025diff}, spatiotemporal inflow generation~\cite{liu2025confild}, and reconstruction of wall-bounded turbulence from sparse measurements~\cite{parikh2026conditional,fan2025generative}. Of particular relevance, the conditional neural field latent diffusion framework CoNFiLD~\cite{du2024conditional} synthesizes full spatiotemporal wall-bounded turbulence while preserving flow inhomogeneity and anisotropy, and supports zero-shot sparse-sensor reconstruction through training-free posterior sampling~\cite{fan2025generative,parikh2026d}. A flow-matching extension of this framework further establishes that patch-local generative priors can recover near-wall small-scale structure with calibrated uncertainty from highly sparse wall measurements~\cite{parikh2026conditional,fan2025generative}, demonstrating the suitability of generative approaches for wall-pressure and wall-velocity reconstruction in turbulent boundary layers. Despite this progress, a fundamental limitation persists in existing generative models of turbulence: they are trained on full-domain fields at a fixed resolution and Reynolds number, coupling the model's representational capacity to its training configuration. A model learned at one domain size or one $Re_\tau$ requires full retraining to generate larger domains or adapt to different flow conditions, a prohibitive requirement in the present setting, where the long-domain, high-$Re_\tau$ wall-pressure data is too expensive to use for training.

Two physical properties of wall turbulence suggest that this coupling between training configuration and generalization capacity is not fundamental, but can be broken by appropriate model design. The first is the spatial locality of high-wavenumber wall-pressure fluctuations: although wall pressure is a nonlocal quantity governed by pressure sources throughout the flow, its high-wavenumber components are dominated by near-wall dynamics and exhibit relatively short correlation lengths in inner units, so a patch-local prior can capture the small-scale physics regardless of the global domain size. The second is the approximate Reynolds-number universality of high-wavenumber statistics in inner scaling: the tail of the wall-pressure spectrum collapses onto a Reynolds-number-independent curve when normalized by $\ell^*$ and the wall shear stress $\tau_w$, as established by experiment~\cite{mcgrath1987some,farabee1991spectral,klewicki2008statistical} and simulation~\cite{choi1990space,yang2022wavenumber,gloerfelt2024aerodynamic,liu2025reynolds}. Together, these properties imply that a patch prior trained in inner-scaled coordinates on short-domain DNS encodes high-wavenumber statistics that generalize across both domain extent and Reynolds number, while the missing low-wavenumber content can be supplied at inference time through sparse sensor conditioning.

To realize this vision, we propose \emph{Patched Flow Matching} (Patched FM): a generative framework that learns a patch-local prior over inner-scaled wall-pressure statistics from short-domain DNS and fuses it with sparse sensor measurements at inference time through training-free posterior sampling. Inspired by patch-based image priors~\cite{hu2024learning}, Patched FM extends this approach to the flow matching setting and, crucially, exploits the physical structure of wall turbulence to achieve generalization across both domain size and Reynolds number. Three original technical contributions underpin the framework.

\emph{(1) Full-spectrum wall-pressure reconstruction from sparse measurements on large domains.}
We formulate and solve a reconstruction problem in which a model trained exclusively on short-domain DNS patches is used to recover the complete wall-pressure spectrum, spanning viscous-scale high-wavenumber content and outer-layer low-wavenumber content simultaneously, on a domain four times larger than the training configuration, from sensor coverage as low as $0.25\%$. This setting is qualitatively distinct from super-resolution (which requires a dense low-resolution input at a fixed upsampling factor), inpainting (which assumes a partially observed field at training resolution), and prior sparse-sensor reconstruction methods (which operate at the training domain size and Reynolds number). The key insight is spectral complementarity: the patch
prior supplies the high-wavenumber content that sensors cannot resolve, while the sensors supply the low-wavenumber content that the patch prior deliberately excludes, with fusion accomplished through training-free posterior sampling at inference time without modifying the learned prior.

\emph{(2) Patch-additive flow matching for domain-size-independent generation.}
We derive a patch-additive decomposition of the marginal vector field in the flow matching setting. By factorizing the full-domain density as a product of patch densities and applying the score-to-velocity identity for linear interpolants, we show that the full-domain ODE velocity decomposes as a sum of independent patch-local contributions, each evaluated by a single shared network. This theoretical result enables both unconditional generation and sensor-conditioned reconstruction on domains of arbitrary size, with patch-parallel ODE integration whose computational cost scales independently of the total domain extent. The shared network incorporates inner-scaled coordinate embeddings---a physics-informed design choice that anchors the learned statistics to viscous scales and is essential for Reynolds-number portability.

\emph{(3) Inner scaling as the physical basis for data-efficient Reynolds-number transfer.}
We establish that expressing the patch prior in inner-scaled coordinates is not merely a normalization convenience but a physical prerequisite for effective Reynolds-number transfer. By encoding statistics in viscous units where the high-wavenumber content is approximately universal, the pre-trained patch prior remains compatible with higher-$Re_\tau$ targets and requires only incremental adaptation rather than wholesale relearning. This
property enables a hierarchical fine-tuning curriculum ($Re_\tau: 180\to500\to1000$) that achieves up to $8\times$ reduction in training cost relative to scratch training, requires only $2.5\%$ of the base training data at each target Reynolds number, and remains effective with as few as $10$ DNS snapshots.
We further show empirically that inner-scaled models trained at one Reynolds number can perform zero-shot reconstruction at a neighboring Reynolds number without any fine-tuning, while outer-scaled models fail under the same conditions---providing direct experimental evidence for the physical argument.

The remainder of the paper is organized as follows. Section~\ref{sec:problem} formalizes the reconstruction task as a Bayesian inverse problem and develops the scale-separation argument. Sections~\ref{sec:fm}--\ref{sec:hie-transfer-learning} present the Patched FM framework. Section~\ref{sec:results}
reports reconstruction results, and Section~\ref{sec:discussion} presents ablation studies and computational cost analysis. Section~\ref{sec:conclusion} summarizes the findings and outlines directions for future work.

\section{Methodology}

\subsection{Problem formulation}
\label{sec:problem}

Wall-pressure fluctuations beneath turbulent boundary layers exhibit a broadband, multi-scale spectral character that reflects the coexistence of physically distinct dynamical regimes. The high-wavenumber tail of the wall-pressure spectrum is set by viscous-scale near-wall dynamics and collapses, to good approximation, onto a Reynolds-number-independent curve when normalized by the viscous length $\ell^* = \nu/u_\tau$ and the wall shear stress $\tau_w=\rho_w u_\tau^2$, as established by both experiment~\cite{farabee1991spectral} and simulation~\cite{liu2025reynolds}. The low-wavenumber content, by contrast, is imprinted by energetic large-scale and very-large-scale motions in the outer layer, whose characteristic streamwise wavelengths scale with the outer length $\delta$ and can reach $\mathcal{O}(10\delta)$ or beyond~\cite{kim1999very,hutchins2007evidence,smits2011high}. Resolving this low-wavenumber content in DNS demands a streamwise domain commensurate with these outer-scale wavelengths, which comes at substantial computational cost. As summarized in Table~\ref{tab:dataset}, extending $L_x$ from $4\pi\delta$ to $16\pi\delta$ at fixed Reynolds number quadruples the streamwise grid; the true cost is compounded further by the extended integration time required to converge low-wavenumber statistics, since the long decorrelation times of the energetic outer-scale motions demand far more snapshots to converge their statistics than the small-scale content. At $Re_\tau = 1000$, for instance, the long-domain case (Case~6) requires a grid of $4800\times 480\times 800$ points, over 1.8 billion in total, and must be integrated for hundreds of bulk time units, $\mathcal{O}(\delta/U_b)$, to achieve statistical convergence. Short-domain DNS, by contrast, resolves the small-scale, high-wavenumber content faithfully but truncates all wavelengths beyond $4\pi\delta$; it cannot, on its own, provide the low-wavenumber statistics needed to characterize the full wall-pressure spectrum.

Sparse wall-mounted pressure sensors constitute a complementary but equally incomplete source of spectral information. Pressure transducer arrays can span streamwise extents far exceeding those of feasible DNS domains, providing access to the energetic low-wavenumber content of the wall-pressure field~\cite{farabee1991spectral,klewicki2008statistical}. However, the inter-sensor spacing is orders of magnitude larger than the viscous length $\ell^*$, so the Nyquist wavenumber of the measurement array falls well below the high-wavenumber range where viscous-scale turbulence resides. Conventional interpolation from sparse sensor data is therefore incapable of recovering fine-scale turbulent structure; the small-scale prior is simply not encoded in the sensor signal, and no post-processing alone can supply it.

Neither source is individually sufficient: short-domain DNS provides the high-wavenumber prior but lacks the domain extent to capture outer-scale low-wavenumber content, while sparse sensors capture the large-scale field but carry no information about the small-scale structure. The present work addresses this complementarity directly. A generative model is trained on an ensemble of short-domain DNS snapshots to learn a patch-local prior over the high-wavenumber wall-pressure statistics; at inference time, sparse long-domain sensor measurements are assimilated as a likelihood constraint to supply the missing low-wavenumber content. Crucially, because the model operates on fixed-size patches in inner-scaled coordinates rather than on full-domain fields, its training configuration is entirely decoupled from the inference-time domain extent, enabling deployment on domains arbitrarily larger than those seen during training.

\begin{figure}[!ht]
    \centering
    \includegraphics[width=1.0\linewidth]{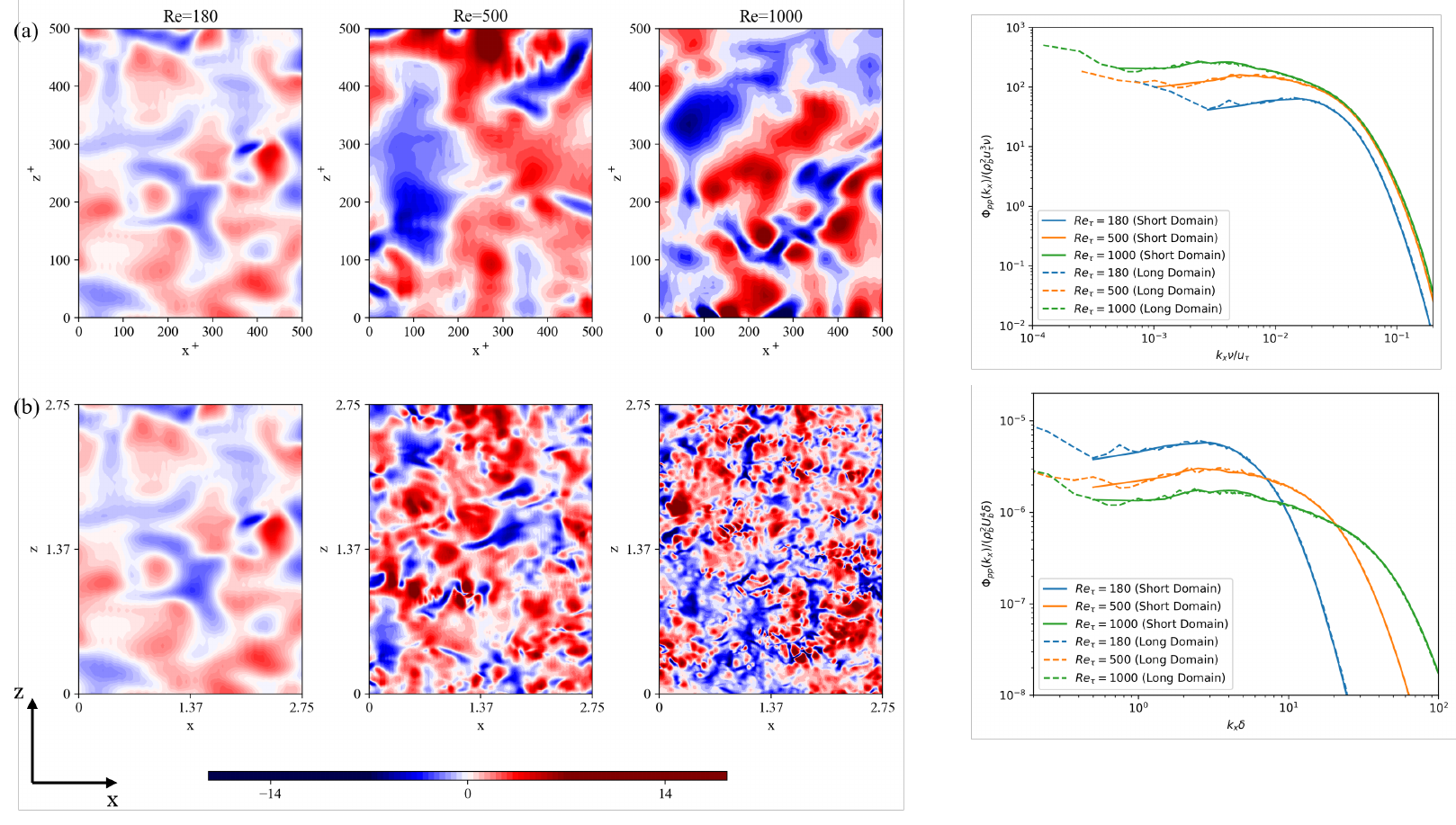}
    \caption{Wall-pressure fluctuations at $Re_\tau = 180$, $500$, and $1000$ visualized in (a) inner-scaled coordinates $(x^+, z^+)$ and (b) outer-scaled coordinates $(x/\delta, z/\delta)$. In inner units, windows of fixed viscous-unit size capture statistically similar high-wavenumber structures across Reynolds numbers, whereas the low-wavenumber content varies systematically with $Re_\tau$ in both coordinate systems. This scale separation motivates the inner-scaled patch decomposition introduced in Section~\ref{sec:patched_fm}.}
    \label{fig:motivation}
\end{figure}

The physical basis for this scale decoupling is illustrated in Figure~\ref{fig:motivation}, which shows wall-pressure snapshots at $Re_\tau = 180$, $500$, and $1000$ in two coordinate systems. In inner-scaled coordinates $(x^+, z^+) = (xu_\tau/\nu,\, zu_\tau/\nu)$ (panel a), windows of fixed viscous-unit dimensions capture small-scale structures of comparable spatial extent and energy density across all three Reynolds numbers, consistent with the approximate inner-layer universality of the high-wavenumber spectrum noted above. In outer-scaled coordinates $(x/\delta, z/\delta)$ (panel b), the same fields differ markedly: as $Re_\tau$ increases, the viscous-scale structures occupy a progressively smaller fraction of $\delta$, highlighting the growing scale separation that makes simultaneous resolution of inner and outer motions difficult. Consequently, a patch-local generative model trained in inner-scaled coordinates encodes high-wavenumber statistics that transfer across Reynolds numbers with minimal retraining, whereas the low-wavenumber content is supplied at inference time through the sparse sensor constraint. This spectral decomposition of responsibilities between the learned patch prior and the data-fidelity term underpins both the patch-based generative framework introduced in Section~\ref{sec:patched_fm} and the hierarchical transfer-learning strategy described in Section~\ref{sec:hie-transfer-learning}.

We now state the problem formally. Let $\phi^S \in \mathbb{R}^{N_x^S \times N_z}$ denote a single wall-pressure snapshot defined on the short domain $\Omega^S = [0, L_x^S] \times [0, L_z]$, and let $\phi^L \in \mathbb{R}^{N_x^L \times N_z}$ denote a snapshot on the long domain $\Omega^L = [0, L_x^L] \times [0, L_z]$, with grid points $N_x^L = 4 N_x^S$, streamwise lengths $L_x^L=4L_x^S=16\pi\delta$, and spanwise length $L_z=4\pi\delta/3$. Here, the superscripts $S$ and $L$ denote short- and long-domain quantities, respectively. The training data consist of an ensemble of short-domain snapshots $\{\phi^S_{(n)}\}_{n=1}^{N_{train}}$. At inference, the goal is to draw samples from the posterior distribution,
\begin{equation}
    \phi^L \sim p(\phi^L \mid \mathbf{y}), \qquad
    \mathbf{y} = \mathcal{F}(\phi^L) + \boldsymbol{\varepsilon}, \qquad
    \boldsymbol{\varepsilon} \sim \mathcal{N}(0, \sigma^2\mathbf{I}),
\end{equation}
where $\mathcal{F}: \mathbb{R}^{N_x^L \times N_z} \to \mathbb{R}^m$ is a linear sparse observation operator that extracts wall-pressure values at $m \ll N_x^L N_z$ fixed sensor locations, and $\boldsymbol{\varepsilon}$ represents measurement noise.

\subsection{Flow matching}
\label{sec:fm}

Flow matching (FM) \cite{lipman2022flow} is a continuous-time generative modeling framework that learns a vector field $v_\theta$ whose induced ODE transports a tractable base distribution $p_0$ to the data distribution $p_1$. Following~\cite{lipman2022flow,parikh2026conditional}, we adopt the linear (affine) interpolant between a Gaussian noise sample $\phi_0 \sim p_0 = \mathcal{N}(0,\mathbf{I})$ and a data sample $\phi_1 \sim p_1 = p_{\mathrm{data}}$:
\begin{equation}
    \phi_s = (1-s)\phi_0 + s\,\phi_1, \qquad s\in[0,1],
    \label{eq:interpolant}
\end{equation}
which defines the conditional vector field $u_s(\phi\mid\phi_1) = \phi_1 - \phi_0$. The associated marginal vector field is the conditional expectation
\begin{equation}
    v_s(\phi) = \mathbb{E}\left[\phi_1 - \phi_0 \middle| \phi_s = \phi\right].
\end{equation}
A neural network $v_\theta(\phi,s)$ is trained to approximate $v_s$ by minimizing the FM regression objective
\begin{equation}
    \mathcal{L}_{\mathrm{FM}}(\theta)
     =
    \mathbb{E}_{s, \phi_0, \phi_1}
    \left\| v_\theta(\phi_s, s) - (\phi_1 - \phi_0) \right\|_2^2,
    \label{eq:fm_loss}
\end{equation}
where the expectation is over $s\sim\mathcal{U}(0,1)$, $\phi_0\sim p_0$, and $\phi_1\sim p_1$. At inference, samples from $p_{\mathrm{data}}$ are obtained by integrating the ODE
\begin{equation}
    \frac{\mathrm{d}\phi_s}{\mathrm{d}s} = v_\theta(\phi_s, s),
    \qquad \phi_0 \sim \mathcal{N}(0,\mathbf{I}),
    \label{eq:ode}
\end{equation}
from $s=0$ to $s=1$.

A key property of the linear interpolant \eqref{eq:interpolant} is that the Tweedie estimator of the clean sample, defined as
$\hat\phi_1(\phi_s,s) = \mathbb{E}[\phi_1\mid\phi_s]$, admits the closed-form expression
\begin{equation}
    \hat\phi_1(\phi_s, s)
     = \phi_s + (1-s)\,v_\theta(\phi_s, s),
    \label{eq:tweedie}
\end{equation}
which provides a direct prediction of the clean field at any integration time $s$ from the current noisy state, and is central to the training-free conditional inference scheme for the patched flow matching framework introduced in Section~\ref{sec:inference}.

\subsection{Patched Flow Matching}
\label{sec:patched_fm}

We introduce Patched FM, a generative framework that learns a patch-local prior over inner-scaled wall-pressure statistics and tiles it across arbitrarily large domains at inference time. The construction adapts the patch-based density factorization of Hu et al.~\cite{hu2024learning}, originally developed for score-based diffusion models, to the flow matching setting, and extends it with inner-scaled coordinate embeddings that enable Reynolds-number transfer.
\begin{figure}[!htp]
    \centering
    \includegraphics[width=1.0\linewidth]{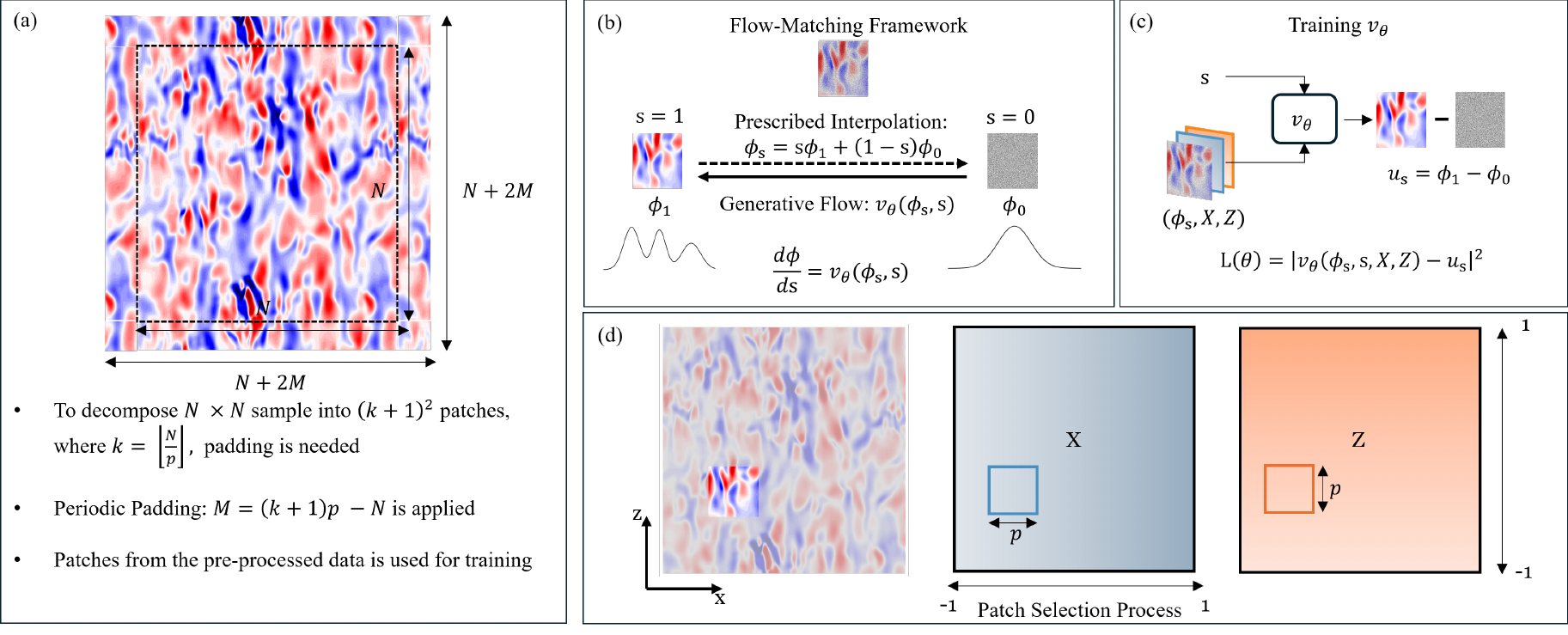}
    \caption{Schematic of the Patched Flow Matching framework. (a) A snapshot of resolution $N\times N$ is extended to $(N+2M)\times(N+2M)$ via periodic padding, where $M=(k+1)P-N$ and $k=\lfloor N/P\rfloor$, and decomposed into $(k+1)^2$ non-overlapping $P\times P$ patches with a randomly chosen tiling origin. (b) The flow matching framework: a linear interpolant
    $\phi_s = s\phi_1 + (1-s)\phi_0$ defines the prescribed path from noise ($s=0$) to data ($s=1$), and the generative ODE $\mathrm{d}\phi/\mathrm{d}s = v_\theta(\phi_s, s)$ transports samples along this path.
    (c) Training: the shared network $v_\theta$ takes a three-channel patch tensor $(\phi_s, X, Z)$ as input and is trained to regress the conditional velocity target $u_s = \phi_1 - \phi_0$.
    (d) Patch selection: patches are drawn at random locations from the inner-scaled wall-pressure field (left); the corresponding normalized coordinate channels $X\in[-1,1]$ and $Z\in[-1,1]$ encode the absolute inner-scaled position of each patch within the domain (center, right).}
    \label{fig:schematic}
\end{figure}

\paragraph{Patch decomposition and tiling}
To avoid boundary artifacts from a naive domain decomposition, a snapshot $\phi$ of resolution $N\times N$ is first extended to
$(N+2M)\times(N+2M)$ via periodic padding, where $k = \lfloor N/P\rfloor$ is the number of patches that fit along each dimension without padding, and $M = (k+1)P - N$ is the additional length needed to accommodate one more patch per dimension (Figure~\ref{fig:schematic}(a)). The padded domain has size $(N+2M) = (k+1)P + M$, which guarantees that any tiling origin $(i,j) \in \{1,\ldots,M\}^2$ yields a valid partition into $(k+1)^2$ non-overlapping $P\times P$ interior patches and a surrounding border region. At each training iteration, a tiling origin $(i,j)$ is drawn uniformly from $\{1,\ldots,M\}^2$, the padded snapshot is partitioned accordingly, and one patch is selected at random from the resulting $(k+1)^2$ patches to update the network. By randomizing the tiling origin across iterations, patch boundaries never coincide with fixed spatial locations, preventing the network from learning decomposition-specific artifacts.

\paragraph{Patch-product density approximation}
The full-domain density $p(\phi)$ is approximated by treating the
patches from each tiling as statistically independent and averaging over all possible tiling origins:
\begin{equation}
    p(\phi) \;\approx\; \frac{1}{\mathcal{Z}}\prod_{i,j=1}^{M}
    \left[p_{i,j,B}(\phi_{i,j,B})
    \prod_{r=1}^{(k+1)^2} p_{i,j,r}(\phi_{i,j,r})\right],
    \label{eq:patch_product}
\end{equation}
where $\phi_{i,j,r}$ and $\phi_{i,j,B}$ denote the $r$-th interior patch and the border region for tiling origin $(i,j)$, and $\mathcal{Z}$ is a normalization constant. The outer product over all origins $(i,j)$ enforces consistency across all possible decompositions of the same field: regardless of where the patch boundaries fall, each patch should be drawn from the same patch distribution. This is an approximation whose validity is tied to the model's intended role. The patch prior is designed to capture only the high-wavenumber, small-scale content of the wall-pressure field. The low-wavenumber content, whose spatial correlations span length scales far exceeding the physical patch extent $P \cdot\Delta x^+$, is supplied at inference time through the sparse sensor constraint (Section~\ref{sec:inference}). The patch independence assumption is therefore only required to hold for the high-wavenumber content, for which spatial correlations are more localized. Its accuracy and the effect of patch size are examined in \ref{app:patch_size}.

\paragraph{Patch-additive velocity field}
Taking the log-gradient of Eq.~\eqref{eq:patch_product} yields the patch-additive score~\cite{hu2024learning}:
\begin{equation}
    \nabla_\phi \log p(\phi) \;=\; \frac{1}{M^2}\sum_{i,j=1}^{M}
    \left[s_{i,j,B}(\phi_{i,j,B})
    + \sum_{r=1}^{(k+1)^2} s_{i,j,r}(\phi_{i,j,r})\right],
    \label{eq:patch_score}
\end{equation}
where $s_{i,j,\cdot} = \nabla_\phi \log p_{i,j,\cdot}$ denotes the score of the corresponding patch density. The key consequence is that the full-domain score decomposes as a sum of independent patch-local contributions.
To transfer this result to the flow matching setting, we use the identity derived in~\cite{parikh2026conditional} (Appendix A.2): for the linear interpolant~\eqref{eq:interpolant}, the marginal vector field $v_s$ at time $s$ is a linear function of the marginal score $\nabla_\phi \log p_s(\phi)$, where $p_s$ is the density of the interpolated field $\phi_s$. Since $p_s$ is obtained by marginalizing $p(\phi)$ through the linear interpolant, the patch-product structure of $p(\phi)$ in Eq.~\eqref{eq:patch_product} propagates to $p_s$, and the patch-additive score~\eqref{eq:patch_score} carries over to the marginal score of $p_s$. The linearity of $v_s$ in the score then yields the patch-additive velocity field:
\begin{equation}
    v_s(\phi) = \frac{1}{M^2}\sum_{i,j=1}^{M}
    \left[v_{s,i,j,B}(\phi_{i,j,B})
    + \sum_{r=1}^{(k+1)^2} v_{s,i,j,r}(\phi_{i,j,r})\right],
    \label{eq:patch_velocity}
\end{equation}
where each patch-local term $v_{s,i,j,r}$ is evaluated by the shared network $v_\theta$. This is the central result: the full-domain marginal vector field decomposes as a sum of patch-local contributions, each computable in parallel by the same network, without any dependence on the global domain extent.

\paragraph{Network input and training}
The wall-pressure data are expressed in inner-scaled coordinates
$(x^+, z^+)$, where $x^+ = xu_\tau/\nu$ and $z^+ = zu_\tau/\nu$. The shared network $v_\theta$ takes as input a three-channel patch tensor $(\phi_s, X, Z)$, where $\phi_s$ contains the wall-pressure values on the patch and $X, Z \in [-1,1]$ are the inner-scaled coordinates linearly normalized to the $[-1,1]$ interval for numerical stability (Figure~\ref{fig:schematic}(c,d)). Because the underlying data are expressed in inner units, these coordinate channels anchor the learned statistics to viscous scales, making the representation approximately portable across Reynolds numbers (Section~\ref{sec:hie-transfer-learning}).
Training minimizes the patch-wise FM objective:
\begin{equation}
    \mathcal{L}(\theta)
    \;=\;
    \mathbb{E}_{s,\,\phi_0,\,\phi_1,\,(i,j),\,r}
    \left\| v_\theta\!\left(\phi_s^{(r)},\, s,\,
    X^{(r)},\, Z^{(r)}\right)
    - \left(\phi_1^{(r)} - \phi_0^{(r)}\right) \right\|_2^2,
    \label{eq:patch_loss}
\end{equation}
where $\phi_s^{(r)} = (1-s)\phi_0^{(r)} + s\phi_1^{(r)}$ is the interpolated patch, and $X^{(r)}, Z^{(r)}$ are the corresponding normalized coordinate arrays. The expectation is over $s \sim \mathcal{U}(0,1)$, noise samples $\phi_0^{(r)} \sim \mathcal{N}(0,
\mathbf{I})$, data patches $\phi_1^{(r)}$ drawn from the short-domain DNS ensemble via the tiling procedure described above, and tiling origins $(i,j)$. Because $v_\theta$ is trained exclusively on fixed-size $P\times P$ patches, it is entirely decoupled from the global domain extent and can be tiled across domains of arbitrary size at inference time.

\begin{figure}[t]
    \centering
    \includegraphics[width=1.0\linewidth]{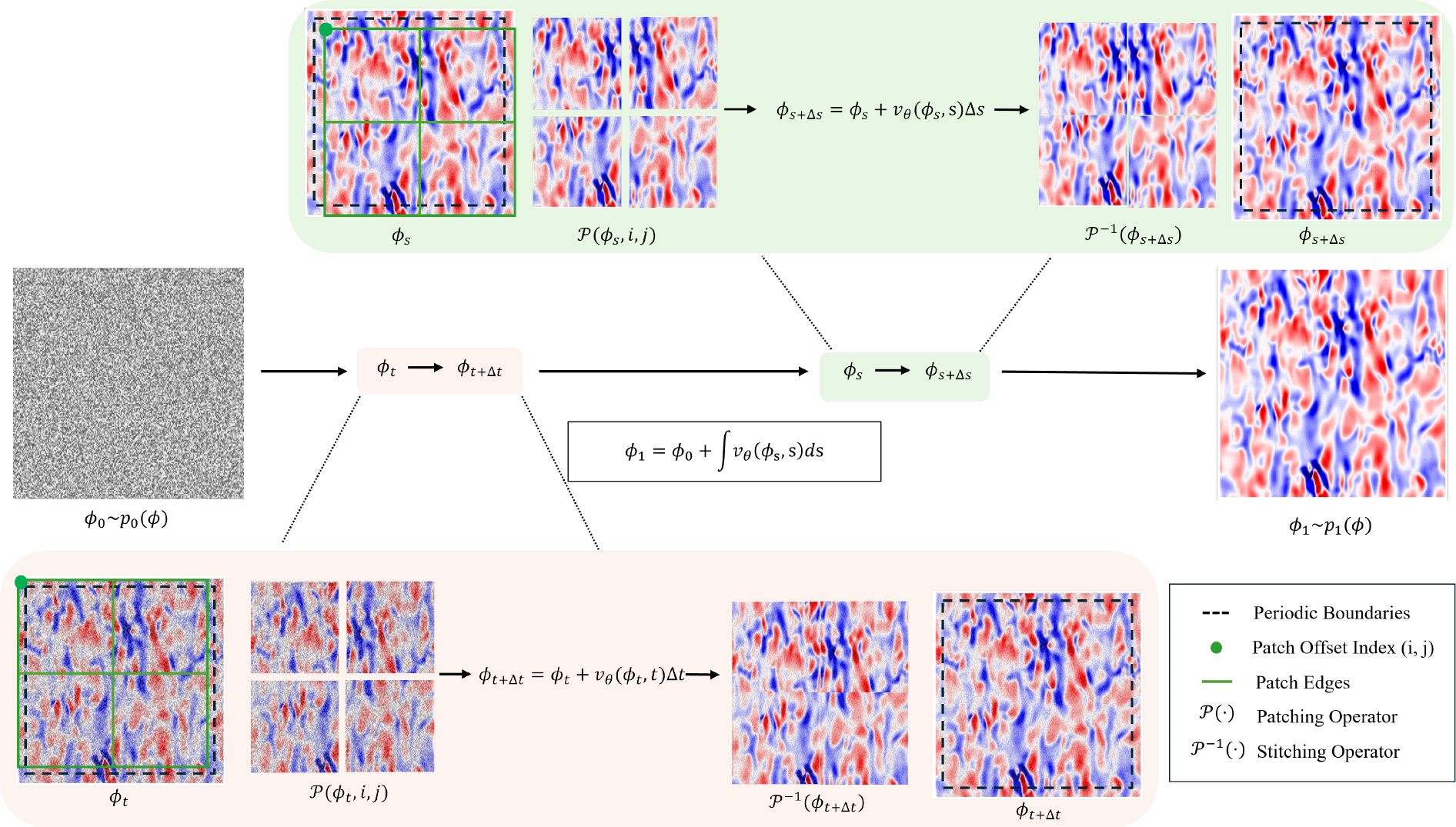}
    \caption{Unconditional inference procedure of Patched FM. At each ODE step, a tiling origin $(i,j)$ is drawn uniformly from $\{1,\ldots,M\}^2$; the patching operator $\mathcal{P}(\cdot,i,j)$ partitions the current field $\phi_s$ into $(k+1)^2$ non-overlapping $P\times P$ patches; each patch is processed in parallel by the shared network $v_\theta$; and the stitching operator $\mathcal{P}^{-1}$ reassembles the updated patches into the full field $\phi_{s+\mathrm{d}s}$. By randomizing the tiling origin at every step, no spatial location is persistently treated as a patch boundary, eliminating boundary artifacts in expectation.}
    \label{fig:unconditional}
\end{figure}
\paragraph{Unconditional generation}
At inference, the target domain is initialized as white Gaussian noise $\phi_0 \sim \mathcal{N}(0,\mathbf{I})$ and the full-domain vector field is approximated at each ODE step by a single randomly sampled tiling, following the Monte Carlo estimator of Eq.~\eqref{eq:patch_velocity}. Specifically, as shown in Figure~\ref{fig:unconditional}, at each step a tiling origin $(i,j)$ is drawn uniformly from $\{1,\ldots,M\}^2$, and the patching operator $\mathcal{P}(\cdot, i, j)$ partitions the current field into $(k+1)^2$ non-overlapping $P\times P$ patches. Each patch is processed in parallel by the shared network $v_\theta$ to obtain a patch-local velocity, and the stitching operator $\mathcal{P}^{-1}$ reassembles the patch velocities into the full-domain vector field. The Euler update then reads
\begin{equation}
    \phi_{s+\Delta s}
    = \phi_s + \mathcal{P}^{-1}\left(
    v_\theta\left(\mathcal{P}(\phi_s, i, j), s, X, Z\right)
    \right)\Delta s,
    \label{eq:euler}
\end{equation}
integrated from $s=0$ to $s=1$ using $100$ uniform steps. By randomizing the tiling origin at every step, the patch boundaries
migrate across the domain throughout the integration, so that no fixed spatial location is persistently treated as a boundary. In expectation over the random origins, this recovers the full patch-additive velocity field of Eq.~\eqref{eq:patch_velocity}, and the resulting generated fields are free of boundary artifacts. Because all patches within a single tiling share one batched network forward pass per ODE step, the wall-clock cost is approximately independent of the total domain size given sufficient parallel hardware.
The patch prior captures the high-wavenumber, small-scale content of the wall-pressure fluctuations; large-scale low-wavenumber features whose wavelengths exceed the physical patch extent $P\cdot\Delta x^+$ are absent from the unconditional prior and must be introduced through the sparse sensor constraint described in Section~\ref{sec:inference}.

\subsection{Training-free conditional generation for long-domain reconstruction}
\label{sec:inference}

The patch prior learned in Section~\ref{sec:patched_fm} captures the high-wavenumber content of the wall-pressure field but carries limited information about the large-scale, low-wavenumber structures, as the patch independence assumption explicitly excludes inter-patch correlations beyond the physical patch extent $P\cdot\Delta x^+$. To reconstruct the full-spectrum wall-pressure field on the long domain, we augment the unconditional ODE with two complementary data-fidelity constraints and generate the long domain autoregressively, one short-domain section at a time.
\begin{figure}[t]
    \centering
    \includegraphics[width=1.0\linewidth]{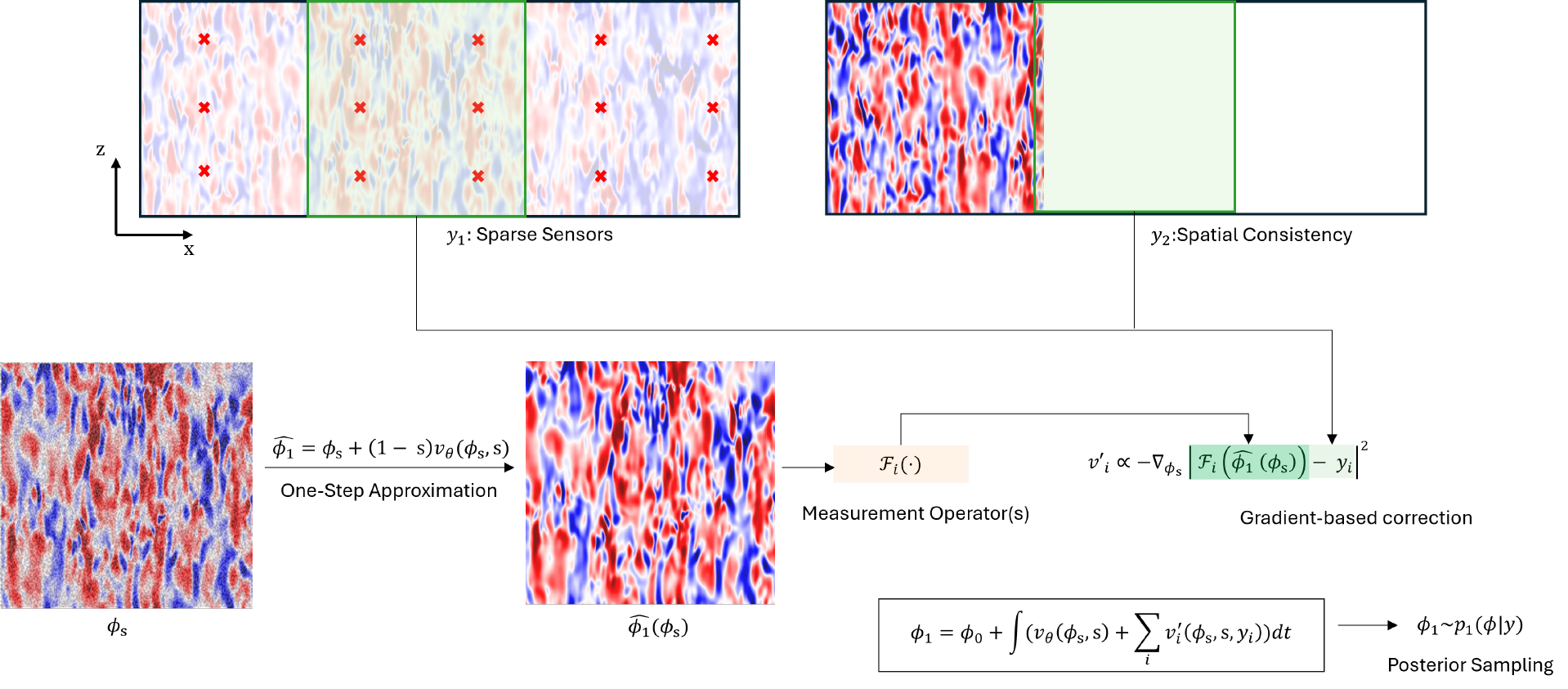}
    \caption{Training-free conditional generation. At each ODE step, the Tweedie estimator produces a one-step prediction $\hat\phi_1(\phi_s)$
    of the clean field from the current noisy state $\phi_s$. Two measurement operators provide complementary constraints: $\mathcal{F}_1$
    extracts sparse sensor observations $y_1$ from the interior of the current section, and $\mathcal{F}_2$ enforces spatial consistency $y_2$
    by matching the left boundary of the current section to the last streamwise column of the previously generated section. The gradient-based corrections $v_1'$ and $v_2'$ from both operators are added to the unconditional velocity $v_\theta$ at each integration step, yielding posterior samples $\phi_1 \sim p_1(\phi \mid \mathbf{y})$.}
    \label{fig:conditional}
\end{figure}

\paragraph{Two-source measurement model}
At inference time, two sources of observational data constrain the generation. The first is a set of sparse wall-pressure measurements
\begin{equation}
    \mathbf{y}_1 = \mathcal{F}_1(\phi^L) + \boldsymbol{\varepsilon}_1,
\end{equation}
where $\mathcal{F}_1: \mathbb{R}^{N_x^L \times N_z} \to \mathbb{R}^{m_1}$ is a sparse subsampling operator that extracts wall-pressure values at $m_1$ sensor locations distributed across the full long domain, and $\boldsymbol{\varepsilon}_1$ represents measurement noise. The second source enforces spatial continuity between autoregressively generated sections. When generating the $\ell$-th section $\phi^{(\ell)}$, the last streamwise column of the previously generated section $\phi^{(\ell-1)}$ serves as a boundary condition:
\begin{equation}
    \mathbf{y}_2^{(\ell)} = \mathcal{F}_2(\phi^{(\ell)}),
    \qquad
    \mathcal{F}_2(\phi^{(\ell)}) = \phi^{(\ell)}|_{x=0},
\end{equation}
where $\mathcal{F}_2$ extracts the left boundary column of the current section, constrained to match the right boundary column of the preceding section. This single-column overlap is sufficient to suppress boundary artifacts between sections while keeping the per-section generation cost independent of the overlap width.

\paragraph{Tweedie-based guidance}
Samples from the posterior $p(\phi^L \mid \mathbf{y}_1, \mathbf{y}_2)$ are obtained without retraining $v_\theta$, by augmenting the unconditional ODE at each integration step with gradient-based corrections from both measurement operators~\cite{chung2022diffusion,parikh2026conditional}. Specifically, the Tweedie estimator~\eqref{eq:tweedie} is applied to the current noisy state $\phi_s$ to obtain a one-step prediction of the clean field. The data-fidelity residuals from both operators are then used to correct the unconditional velocity:
\begin{equation}
    \tilde{v}_\theta(\phi_s, s) = v_\theta(\phi_s, s) + \sum_{i=1}^{2} v_i'(\phi_s, s, \mathbf{y}_i),
    \label{eq:guided_full}
\end{equation}
where each correction term is
\begin{equation}
    v_i'(\phi_s, s, \mathbf{y}_i) =  -\gamma_s^{(i)}\,
    \nabla_{\phi_s}
    \left\|\mathbf{y}_i
    - \mathcal{F}_i\left(\hat\phi_1(\phi_s,s)\right)
    \right\|_2^2,
    \label{eq:correction}
\end{equation}
with the guidance weight
\begin{equation}
    \gamma_s^{(i)} =
    \frac{\left\|v_\theta(\phi_s,s)\right\|}
         {\left\|\mathbf{y}_i
         - \mathcal{F}_i\left(\hat\phi_1(\phi_s,s)\right)
         \right\|},
    \label{eq:gamma}
\end{equation}
normalizing each likelihood gradient by the current residual magnitude to dynamically balance the generative prior and the measurement likelihood throughout integration. The gradient $\nabla_{\phi_s}$ in Eq.~\eqref{eq:correction} is evaluated by automatic differentiation through the Tweedie estimator and the observation operator $\mathcal{F}_i$.

\paragraph{Spatial autoregressive generation}
Since the long domain is $N_l$-times the training domain, direct generation of the full domain in a single pass would require the patch prior to supply low-wavenumber coherence across the entire extent $L_x^L$---a scale far beyond what the patch prior can represent. Instead, we generate the long domain autoregressively as $N_l$ consecutive sections, each of size $[0, 4\pi\delta]\times[0, 4\pi\delta/3]$, matching the training configuration. This design provides a complementary pathway for recovering low-wavenumber content: the boundary constraint $\mathbf{y}_2^{(\ell)}$ propagates phase and amplitude information about large-scale outer-layer motions from one section to the next, enabling coherent structures whose streamwise extent exceeds $4\pi\delta$ to be reconstructed progressively. The two constraints are thus complementary: $\mathbf{y}_1$ supplies global low-wavenumber information distributed across the full domain, while $\mathbf{y}_2$ enforces local streamwise coherence at section interfaces.

For the first section ($\ell=1$), only the sparse sensor constraint $\mathbf{y}_1$ is active. For each subsequent section ($\ell = 2, \ldots, N_l$), both constraints are active: $\mathbf{y}_1$ provides the sparse sensor measurements within the current section's spatial extent, and $\mathbf{y}_2^{(\ell)}$ pins the left boundary of the current section to the last streamwise column of the previously generated section $\phi^{(\ell-1)}$, ensuring seamless spatial continuity. The full conditional ODE for each section is integrated from $s=0$ to $s=1$ using $100$ Euler steps:
\begin{equation}
    \phi_1^{(\ell)}
    = \phi_0 + \int_0^1
    \left(v_\theta(\phi_s, s)
    + \sum_{i} v_i'(\phi_s, s, \mathbf{y}_i^{(\ell)})\right)
    \mathrm{d}s.
    \label{eq:conditional_ode}
\end{equation}
The $N_l$ generated sections are concatenated to form the complete long-domain reconstruction $\phi^L$. Because $v_\theta$ remains strictly patch-local, only the data-fidelity gradients $v_i'$ couple patches through the global operators $\mathcal{F}_i$, requiring a single global gather/scatter per ODE step, and the patch-parallel structure of the unconditional generation is preserved throughout.

\subsection{Hierarchical transfer learning across $Re_\tau$}
In this work the patch prior is intended to focus on the high-wavenumber, inner-scaled component of the wall-pressure field. As discussed above, this component exhibits approximate Reynolds-number universality when expressed in viscous units, whereas the low-wavenumber outer-layer content varies substantially with $Re_\tau$ and is supplied at inference time through sparse sensor conditioning. Consequently, the transferability of Patched FM across Reynolds numbers depends primarily on the extent to which the learned patch prior captures these approximately universal high-wavenumber statistics. This observation motivates a hierarchical transfer-learning strategy. Rather than training an independent model at each Reynolds number, we sequentially adapt a model trained at lower $Re_\tau$ to progressively higher Reynolds numbers. Specifically, a model first trained at $Re_\tau=180$ is fine-tuned at $Re_\tau=500$, and the resulting model is subsequently adapted to $Re_\tau=1000$. The rationale follows directly from the inner-scaled spectral behavior shown in Figure~\ref{fig:motivation}: the high-wavenumber portions of the wall-pressure spectra exhibit increasing collapse with Reynolds number, with the $Re_\tau=500$ and $1000$ spectra nearly indistinguishable over a substantial range of wavenumbers. Since the patch prior is designed to model this portion of the spectrum, only incremental adaptation is expected to be necessary as Reynolds number increases.

This hierarchical curriculum provides two advantages. First, it reduces the amount of high-$Re_\tau$ training data required to obtain an accurate prior, since much of the small-scale statistical structure has already been learned at lower Reynolds numbers. Second, it preserves the physical interpretation of the patch prior as an approximately universal representation of inner-layer wall-pressure statistics, while allowing limited adaptation to Reynolds-number-dependent effects that remain present in finite-Reynolds-number flows.

To further improve data efficiency and reduce overfitting, we employ parameter-efficient fine-tuning. Only the final layers of the UNet backbone are updated during transfer, while the majority of the pre-trained network remains frozen. Specifically, we fine-tune the three ResBlocks operating at the original spatial resolution together with the final convolutional projection layer (see \ref{app:unet} for details of the network). These layers comprise approximately $7\%$ of the total trainable parameters. This choice provides sufficient flexibility to adapt the high-wavenumber statistics across Reynolds numbers while retaining the previously learned patch representations and mitigating overfitting to the limited high-$Re_\tau$ training data.

\section{Numerical results}
\label{sec:results}

We evaluate the proposed Patched FM framework across three friction Reynolds numbers, $Re_\tau = 180$, $500$, and $1000$, in two operational modes. The first is unconditional generation on the short training domain ($L_x^S = 4\pi\delta$), which probes the fidelity of the learned patch prior against short-domain DNS. The second, and primary, mode is conditional reconstruction on the long domain ($L_x^L = 16\pi\delta$), where the model assimilates sparse sensor measurements to recover the full wall-pressure spectrum including the low-wavenumber content absent from the training data; the ground truth for all conditional experiments is long-domain DNS, which the model never accesses during training. These two modes together demonstrate two distinct levels of generalization: full-domain spatial generation from patch-scale training, and domain-size extrapolation from the $4\pi\delta$ training extent to the $16\pi\delta$ testing extent.

\subsection{Experimental setup}
\label{sec:setup}

\paragraph{DNS dataset}
All simulations are compressible channel flows at bulk Mach number $M_b = 0.4$, obtained using a high-order non-dissipative finite-difference scheme with sixth-order spatial discretization and fourth-order Runge--Kutta time integration, which resolves both the hydrodynamic and acoustic contributions to the wall-pressure fluctuations with high fidelity~\cite{liu2024subconvective,liu2025reynolds}. The dataset covers three friction Reynolds numbers, $Re_\tau = 180$, $500$, and $1000$, each at two streamwise domain lengths, yielding six cases summarized in Table~\ref{tab:dataset}. Short-domain cases (Cases~1--3, $L_x^S = 4\pi\delta$) provide the training data; long-domain cases (Cases~4--6, $L_x^L = 16\pi\delta$) serve exclusively as the test ground truth for conditional reconstruction. Wall-pressure snapshots from both walls are used, doubling the effective snapshot count at no additional simulation cost. Further details are given in~\cite{liu2024subconvective,liu2025reynolds}.
\begin{table}[t]
    \caption{Domain and grid sizes of the channel flow DNS dataset.
    Cases 1--3 (short domain, $L_x^S = 4\pi\delta$) are used for
    training; Cases 4--6 (long domain, $L_x^L = 16\pi\delta$) are
    reserved exclusively for testing.}
    \label{tab:dataset}
    \centering
    \begin{tabular}{cccccc}
        \hline
        Case & $Re_\tau$ & Domain & $L_x\times L_y\times L_z$
             & $N_x\times N_y\times N_z$ & Training/Testing\\
        \hline
        1 & 180  & Short & $4\pi\delta\times 2\delta\times 4\pi\delta/3$
          & $256\times 192\times 256$ & Training \\
        2 & 500  & Short & $4\pi\delta\times 2\delta\times 4\pi\delta/3$
          & $600\times 320\times 400$ & Training\\
        3 & 1000 & Short & $4\pi\delta\times 2\delta\times 4\pi\delta/3$
          & $1200\times 480\times 800$ & Training\\
        \hline
        4 & 180  & Long  & $16\pi\delta\times 2\delta\times 4\pi\delta/3$
          & $1024\times 192\times 256$ & Testing\\
        5 & 500  & Long  & $16\pi\delta\times 2\delta\times 4\pi\delta/3$
          & $2400\times 320\times 400$ & Testing\\
        6 & 1000 & Long  & $16\pi\delta\times 2\delta\times 4\pi\delta/3$
          & $4800\times 480\times 800$ & Testing\\
        \hline
    \end{tabular}
\end{table}

\paragraph{Training configuration}
The Patched FM model is trained on a stochastic mixture of patch sizes following the variable-patch-size scheme of~\cite{hu2024learning}, drawing sizes of $16$, $32$, and $56$ grid points with sampling probabilities $0.2$, $0.3$, and $0.5$, respectively; the largest patch, $P = 56$, sets the physical extent $P\cdot\Delta x^+$ that captures the high-wavenumber, viscous-scale content of the wall-pressure field while remaining small enough to satisfy the patch-independence assumption of Section~\ref{sec:patched_fm}. The base model is trained on Case~1 ($Re_\tau = 180$) using $20{,}000$ snapshots. Models at $Re_\tau = 500$ and $1000$ are obtained by hierarchical transfer learning (Section~\ref{sec:hie-transfer-learning}), fine-tuning sequentially $180 \to 500 \to 1000$ using only $500$
short-domain snapshots per Reynolds number (Cases~2 and~3), corresponding to $2.5\%$ of the base training set. Random patch sampling over the available snapshots produces an effective training set substantially larger than the snapshot count, which is key to enabling generalization within the patch decomposition framework.

\paragraph{Evaluation}
Unconditional generation quality is assessed via one-dimensional streamwise and spanwise wavenumber spectra of wall-pressure fluctuations, $\Phi_{pp}(k_x)$ and $\Phi_{pp}(k_z)$, compared against the short-domain DNS reference. Conditional reconstruction quality is assessed via the same spectral diagnostics against the long-domain DNS ground truth, with cubic interpolation from the sparse sensor locations as a deterministic baseline.

\subsection{Generation at \texorpdfstring{$Re_\tau = 180$}{Re=180}}
\label{sec:re180}

The $Re_\tau = 180$ case establishes the baseline performance of the Patched FM framework before any Reynolds-number transfer is applied. The model is trained on short-domain DNS (Case~1) and evaluated in both unconditional and conditional modes.

\begin{figure}[!ht]
    \centering
    \includegraphics[width=0.8\linewidth]{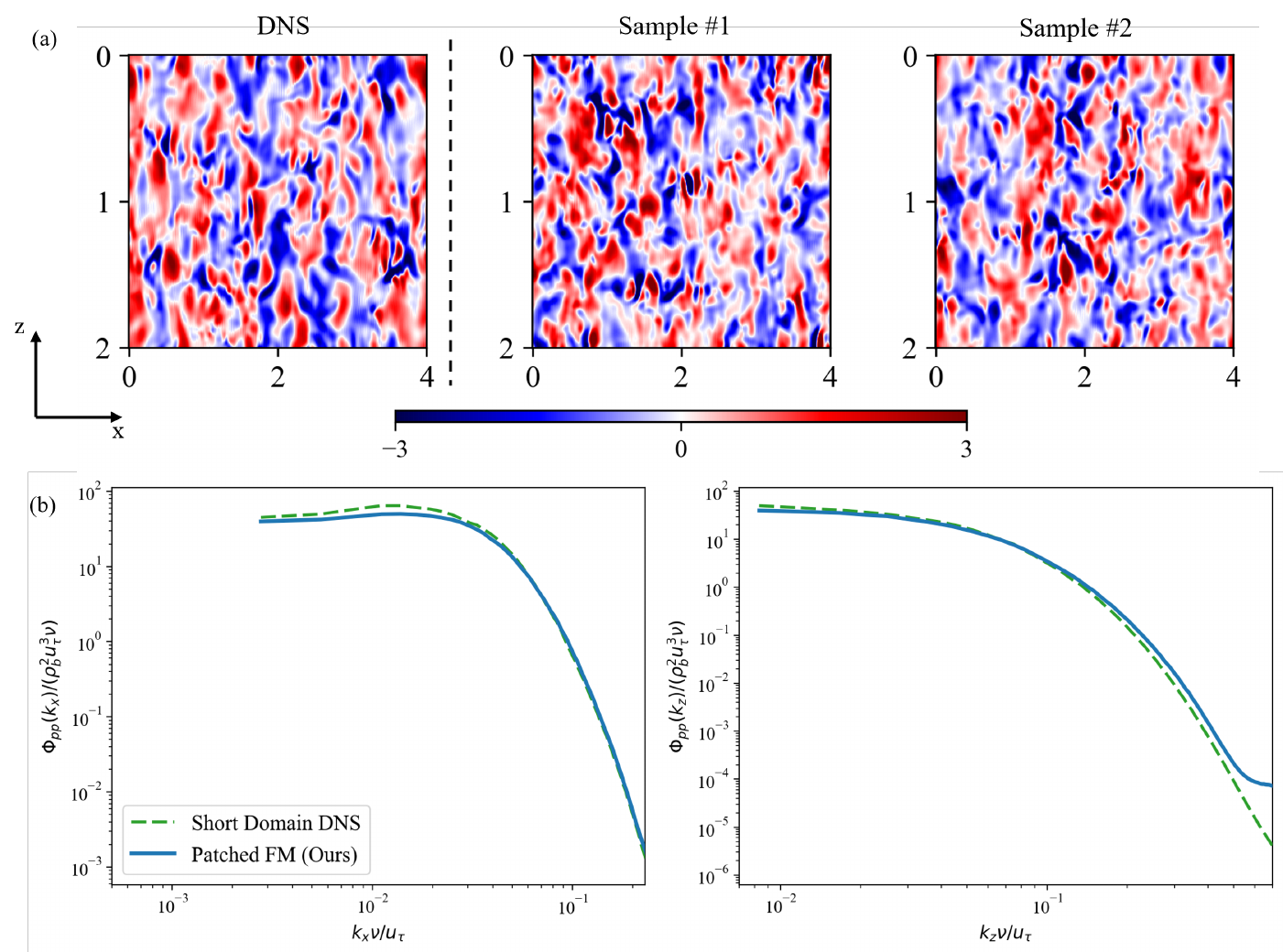}
    \caption{Unconditional generation at $Re_\tau=180$: (a) Contours of the wall-pressure fluctuations from DNS and two independently generated samples on the short training domain ($L_x^S = 4\pi\delta$); (b) streamwise and spanwise wavenumber spectra $\Phi_{pp}(k_x)$ and $\Phi_{pp}(k_z)$ from an ensemble of $500$ generated samples, compared with short-domain DNS.}
    \label{fig:re180_uncond}
\end{figure}

\paragraph{Unconditional generation}
Figure~\ref{fig:re180_uncond}(a) shows two independently generated samples alongside a representative DNS snapshot on the training domain. The generated fields reproduce the spatial organization and multiscale distribution of wall-pressure fluctuations characteristic of $Re_\tau = 180$, with
no visible discontinuities at patch boundaries. To quantify statistical fidelity, Figure~\ref{fig:re180_uncond}(b) compares the streamwise and spanwise wavenumber spectra, calculated from an ensemble of $500$ generated samples, against the short-domain DNS reference. The Patched FM spectra closely match DNS across the full resolved wavenumber range in both directions, confirming that the model accurately reproduces the energy-containing scales and the dissipative tail of the wall-pressure spectrum.
Two aspects of this result are worth emphasizing. First, the network operates exclusively on fixed-size $P\times P$ patches at both training and inference time; the globally coherent full-domain fields emerge from the patch-additive velocity field of Eq.~\eqref{eq:patch_velocity} combined with the randomized tiling strategy, which prevents any fixed spatial location from being persistently treated as a patch boundary. Second, the training domain ($L_x^S = 4\pi\delta$) is itself larger than the patch extent $P\cdot\Delta x^+$, so unconditional generation on the training domain already constitutes a non-trivial spatial extrapolation beyond the patch scale. The ability of the model to reproduce the full short-domain spectrum without artifact validates the patch-additive decomposition as a faithful approximation to the full-domain generative process, and provides the foundation for the more demanding domain-size extrapolation demonstrated in the conditional experiments below.

\paragraph{Conditional reconstruction on the long domain}
We now verify one of the central hypotheses of this work: that a model trained exclusively on short-domain patches can generalize to long-domain reconstruction by leveraging the training-free conditional inference strategy of Section~\ref{sec:inference}. The model is conditioned on sparse sensor measurements drawn from the long-domain DNS at $Re_\tau = 180$ (Case~4, $L_x^L = 16\pi\delta$)---a dataset the model never accessed during training---with only $1\%$ of the long-domain grid points observed. Conditional inference proceeds autoregressively through four consecutive sections, each constrained by the sparse sensors within its spatial extent and by the right boundary of the previously generated section, as described in Section~\ref{sec:inference}.

\begin{figure}[!ht]
    \centering
    \includegraphics[width=1.0\linewidth]{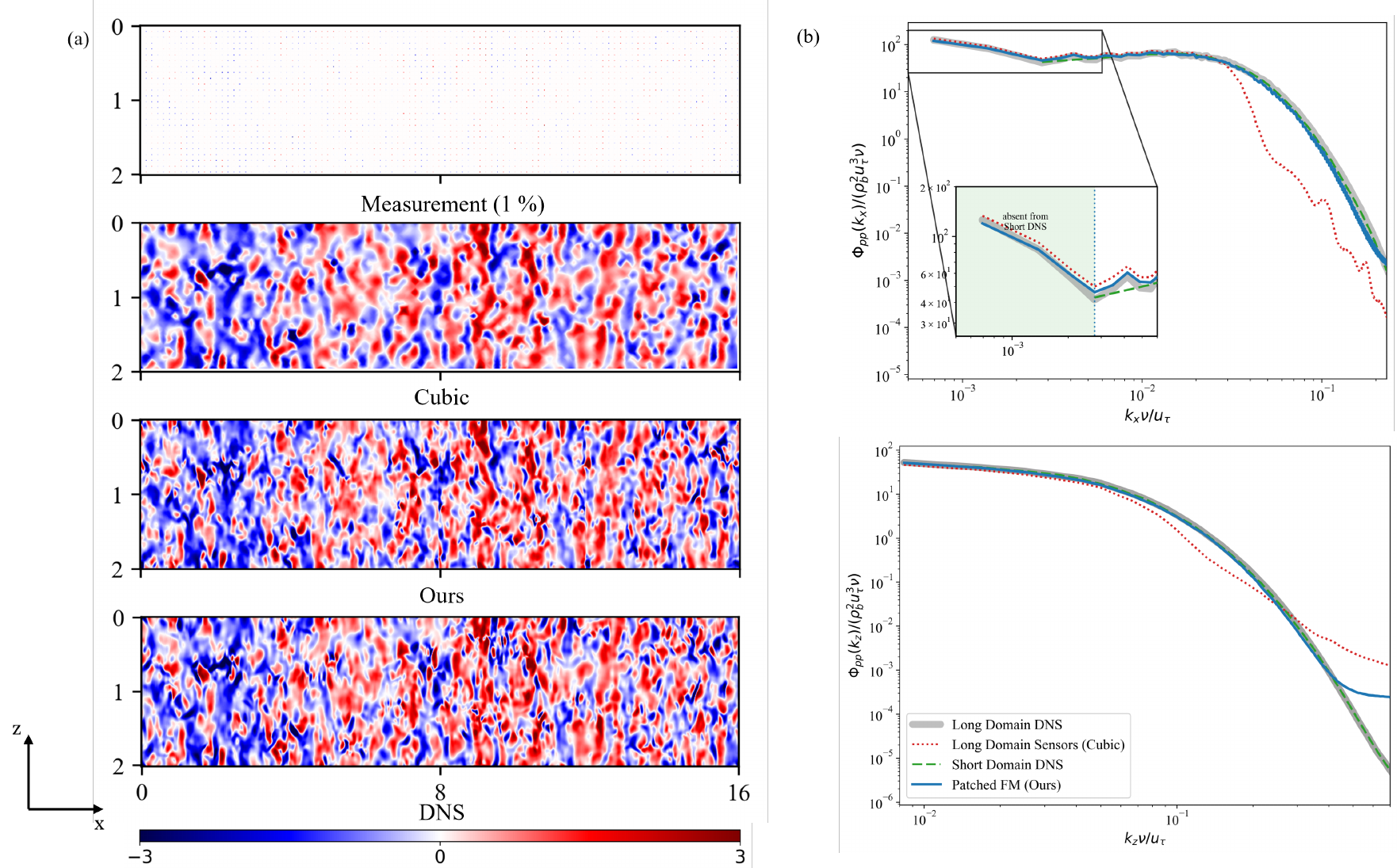}
    \caption{Conditional reconstruction at $Re_\tau=180$ on the long
    domain ($L_x^L = 16\pi\delta$, Case~4): (a) Contours of the wall-pressure fluctuations
    from sparse sensor measurements ($1\%$ coverage), cubic interpolation,
    Patched FM reconstruction, and long-domain DNS ground truth; (b)
    streamwise and spanwise wavenumber spectra $\Phi_{pp}(k_x)$ and $\Phi_{pp}(k_z)$, comparing Patched FM,
    cubic interpolation, short-domain DNS, and long-domain DNS.}
    \label{fig:re180_cond}
\end{figure}
Figure~\ref{fig:re180_cond}(a) compares the Patched FM reconstruction and cubic interpolation against the long-domain DNS ground truth. Cubic interpolation captures only the coarsest large-scale features and introduces severe smoothing at all smaller scales. The Patched FM reconstruction, by contrast, recovers the large-scale organization imposed by the sparse sensor constraint while simultaneously populating the field with physically consistent small-scale turbulent structures that cubic interpolation cannot supply.
For a statistically robust assessment, $500$ independent reconstructions are generated, each conditioned on a distinct sparse-sensor realization drawn from the same long-domain DNS, probing the framework's consistency across different measurement configurations. The spectral comparison in
Figure~\ref{fig:re180_cond}(b) quantifies the reconstruction fidelity. In the streamwise direction, the Patched FM ensemble closely matches the long-domain DNS spectrum across all resolved wavenumbers, faithfully recovering the elevated low-wavenumber energy that is entirely absent from the short-domain DNS due to its truncated streamwise extent. In the spanwise direction, the Patched FM spectrum overlaps with the DNS reference across virtually all resolved scales, with only a minor deviation at the highest wavenumbers. Cubic interpolation, by contrast, deviates substantially from DNS at intermediate and high
wavenumbers in both directions, confirming that recovering the multi-scale structures of the wall-pressure field from sparse measurements requires the high-wavenumber generative prior that Patched FM provides.

These results confirm that the patch-additive generative prior, when combined with training-free sensor conditioning, can reconstruct statistically faithful wall-pressure fields on a domain four times larger than the training configuration. The physical basis for this generalization is the approximate
Reynolds-number invariance of inner-scaled wall-pressure statistics at the patch scale: the small-scale prior learned from short-domain patches transfers directly to the long domain, while the sparse sensors supply the missing low-wavenumber content.

\subsection{Hierarchical transfer learning for \texorpdfstring{$Re_{\tau}$}{Re}= 500 and 1000}
\label{sec:hie-transfer-learning}

The approximate Reynolds-number invariance of inner-scaled high-wavenumber wall-pressure statistics motivates our second hypothesis: that the $Re_\tau = 180$ model can be extended to higher Reynolds numbers through hierarchical transfer learning, at a fraction of the data cost required for training from scratch. The base model is fine-tuned sequentially $180 \to 500 \to 1000$, using the short-domain DNS at each target Reynolds number (Cases~2 and~3). Critically, only $500$ snapshots per Reynolds number are required ($2.5\%$ of the $20{,}000$ snapshots used at $Re_\tau = 180$), because the high-wavenumber content encoded in the pre-trained weights is largely shared across Reynolds numbers in inner scaling, so the network needs only to adapt to
the $Re_\tau$-dependent modifications rather than relearn the universal small-scale statistics from scratch. As we demonstrate in next subsection, this data efficiency extends further: transfer learning remains effective with as few as $10$ DNS snapshots, eliminating the data-acquisition bottleneck that typically limits the applicability of generative models to high-Reynolds-number turbulence.

\begin{figure}[!ht]
    \centering
    \includegraphics[width=0.8\linewidth]{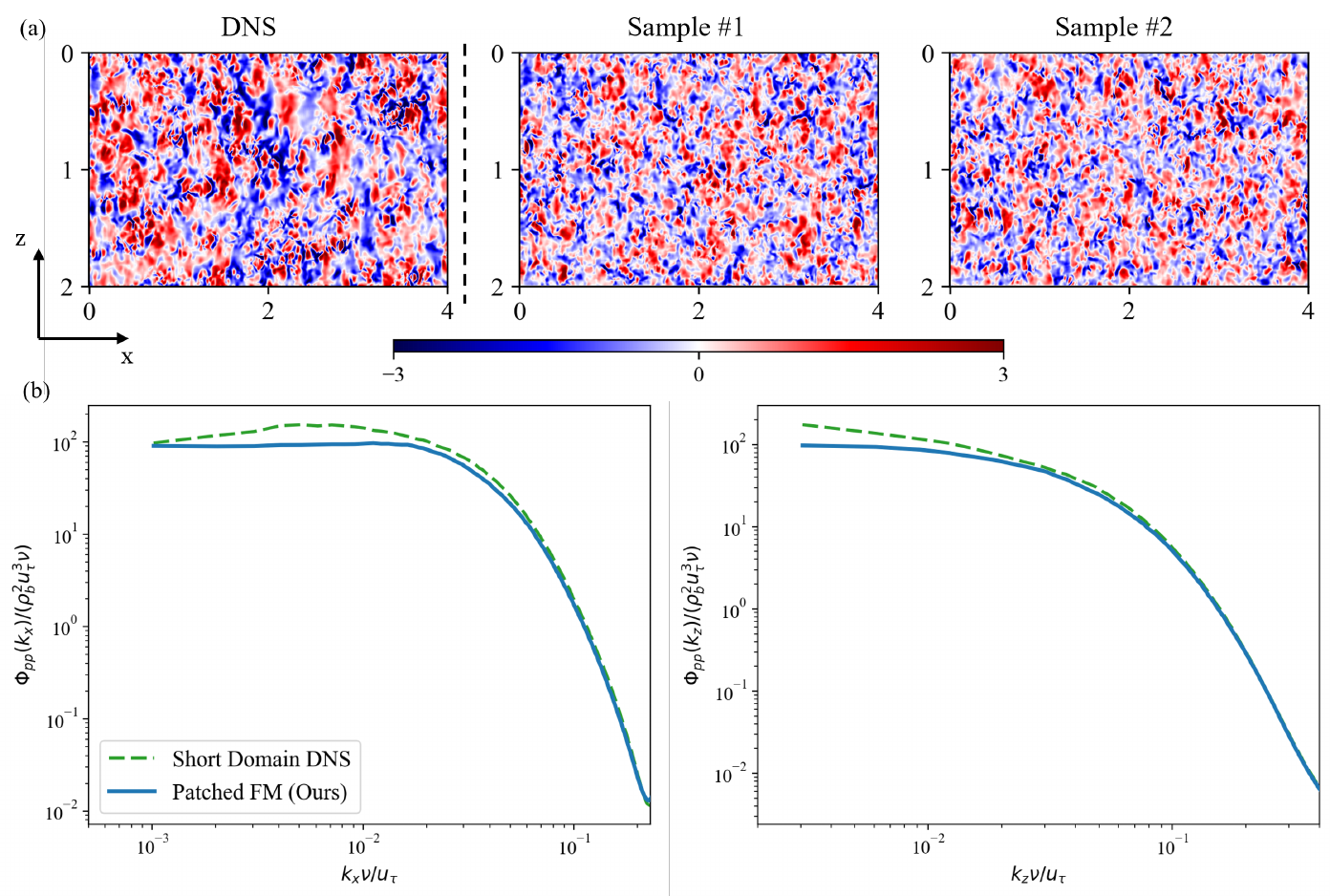}
    \caption{Unconditional generation at $Re_\tau=500$ after hierarchical transfer from $Re_\tau=180$: (a) Contours of the wall-pressure fluctuations from DNS and two generated samples on the short training domain ($L_x^S = 4\pi\delta$); (b) streamwise and spanwise wavenumber spectra $\Phi_{pp}(k_x)$ and $\Phi_{pp}(k_z)$ from $100$ generated samples, compared with DNS.}
    \label{fig:re500_uncond}
\end{figure}

\paragraph{Results at $Re_\tau = 500$}
Figure~\ref{fig:re500_uncond} shows the unconditional generation
performance after fine-tuning to $Re_\tau = 500$. The generated
wall-pressure fluctuation fields reproduce the heightened intermittency and finer structures characteristic of the higher Reynolds number. The wall-pressure spectra from $100$ generated samples agree well with DNS across most of the resolved range, with a detectable under-estimation at the lowest wavenumbers. This low-wavenumber deficit is a direct consequence of the patch-independence approximation: at $Re_\tau = 500$, a fixed inner-unit patch of $56\times 56$ grid points covers a smaller fraction of $\delta$ than at $Re_\tau = 180$ (Figure~\ref{fig:motivation}),
so wavelengths exceeding the physical patch extent $P\cdot\Delta x^+$ fall outside the support of the unconditional prior. This limitation is examined in \ref{app:patch_size} and is corrected by the conditional inference pipeline, as shown next.

\begin{figure}[!ht]
    \centering
    \includegraphics[width=1.0\linewidth]{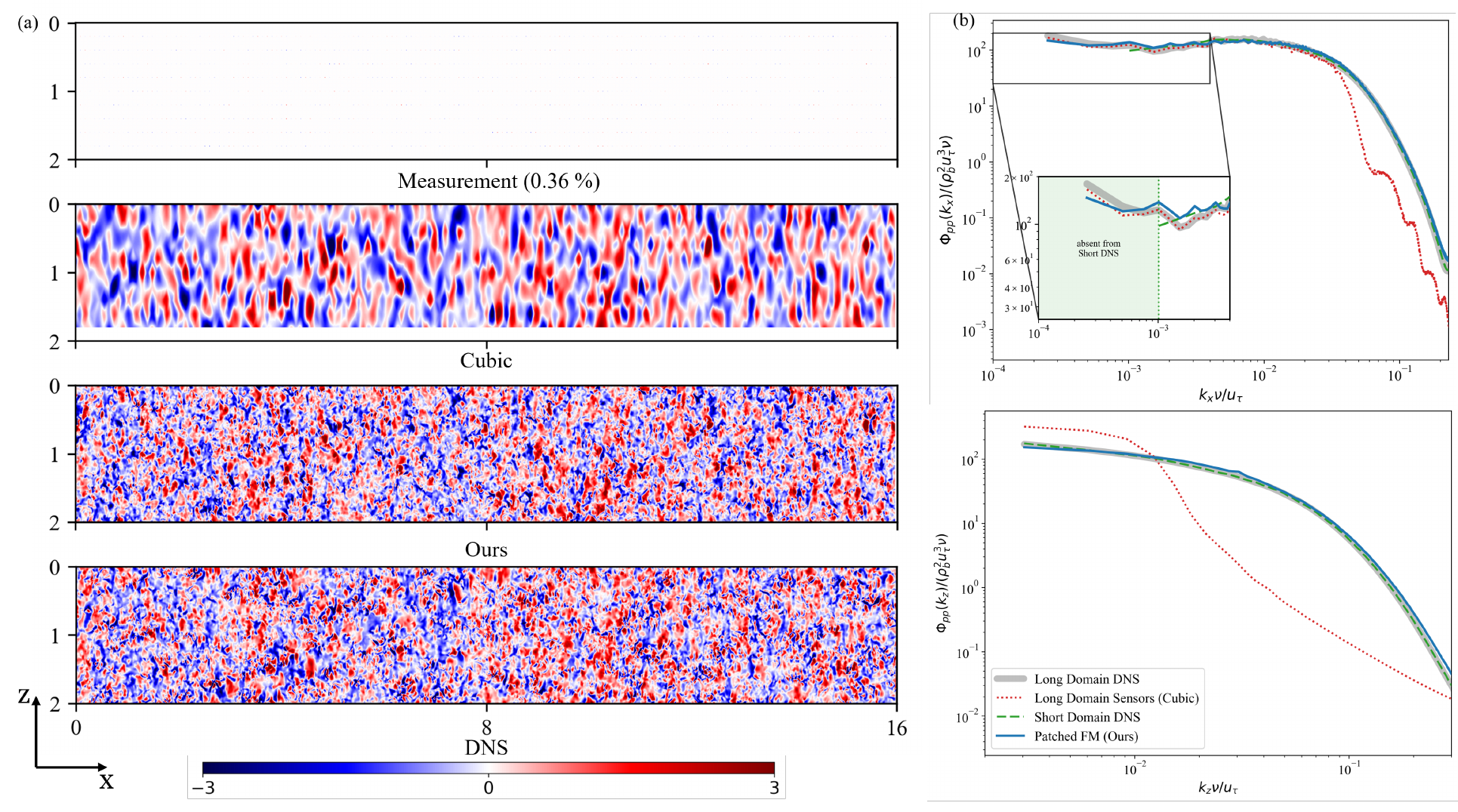}
    \caption{Conditional reconstruction at $Re_\tau=500$ on the long
    domain ($L_x^L = 16\pi\delta$, Case~5): (a) Contours of the wall-pressure fluctuations from sparse sensor measurements ($0.36\%$ coverage), cubic interpolation, Patched FM reconstruction, and long-domain DNS; (b) streamwise and spanwise wavenumber spectra $\Phi_{pp}(k_x)$ and $\Phi_{pp}(k_z)$, comparing Patched FM, cubic interpolation, short-domain DNS, and long-domain DNS.}
    \label{fig:re500_cond}
\end{figure}

For conditional reconstruction on the long domain (Case~5, $0.36\%$ sensor coverage), Figure~\ref{fig:re500_cond} demonstrates that the sparse-sensor likelihood restores the low-wavenumber content that the unconditional prior under-represents. The Patched FM ensemble
of $100$ reconstructions closely follows the long-domain DNS spectrum in both directions, including the low-wavenumber streamwise upturn absent from the short-domain reference. Cubic interpolation produces a severely over-smoothed field and fails to match DNS at any but the lowest wavenumbers, confirming that recovering the multi-scale wall-pressure structure at $Re_\tau = 500$ requires the high-wavenumber prior that only the generative model can supply.

\begin{figure}[!ht]
    \centering
    \includegraphics[width=0.8\linewidth]{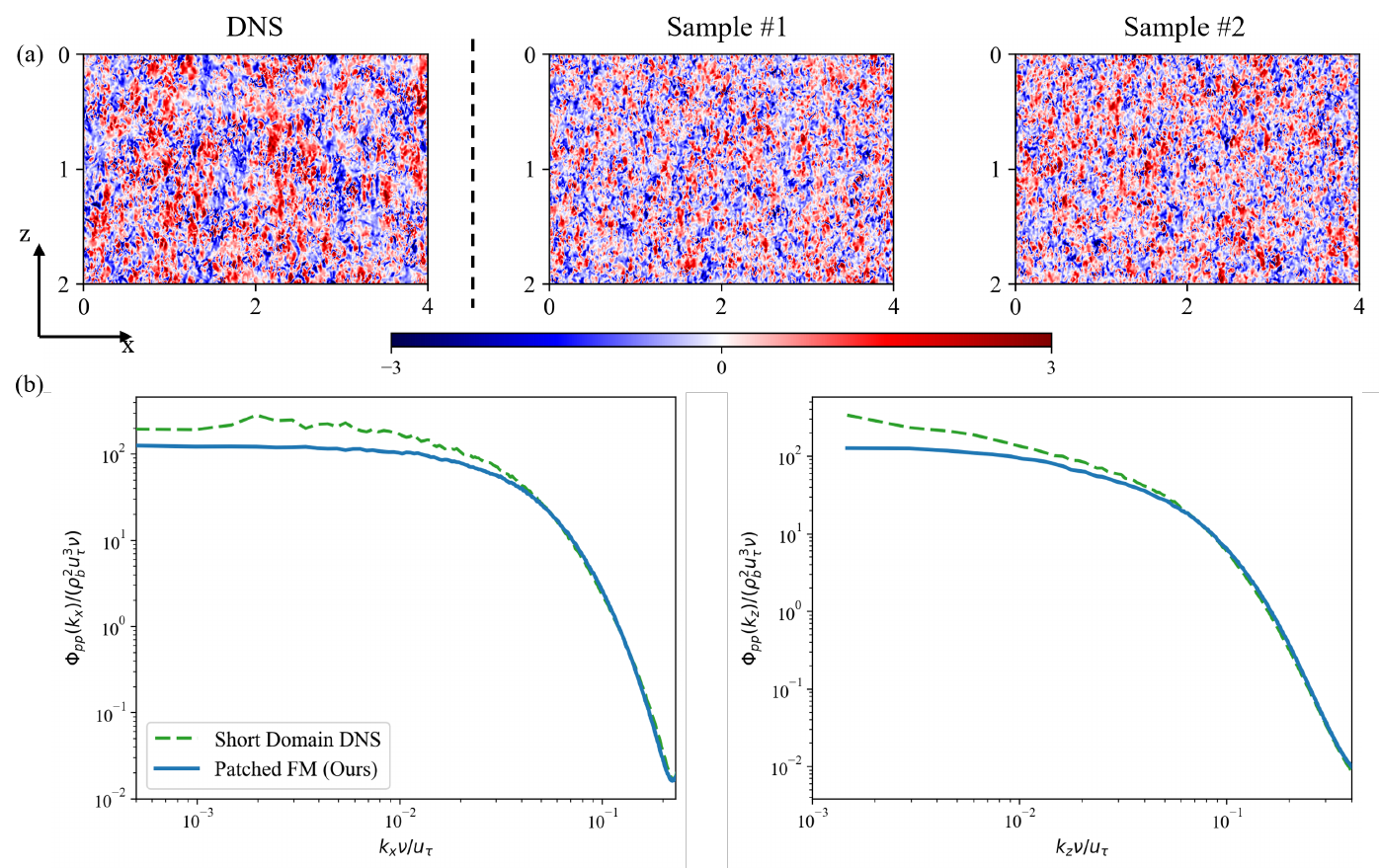}
    \caption{Unconditional generation at $Re_\tau=1000$ after hierarchical transfer from $Re_\tau=500$: (a) Contours of the wall-pressure fluctuations from DNS and two generated samples on the short training domain ($L_x^S = 4\pi\delta$); (b) streamwise and spanwise wavenumber spectra $\Phi_{pp}(k_x)$ and $\Phi_{pp}(k_z)$ from $100$ generated samples, compared with DNS.}
    \label{fig:re1000_uncond}
\end{figure}

\paragraph{Results at $Re_\tau = 1000$}
The model fine-tuned to $Re_\tau = 500$ is further adapted to $Re_\tau = 1000$ using Case~3. The cost advantage of this ascending-order hierarchical strategy over direct transfer from $Re_\tau = 180$ or training from scratch is quantified in Section~\ref{sec:train_and_inf_costs}.
Figure~\ref{fig:re1000_uncond} shows the unconditional generation results. The generated fields display the expected increase in small-scale content and heightened intermittency relative to $Re_\tau = 500$. The low-wavenumber deficit in the unconditional spectra is wider than at $Re_\tau = 500$, consistent with the shrinking ratio of patch extent to $\delta$ as $Re_\tau$ increases: the $56\times 56$-point patch covers a progressively smaller outer-scale extent, so an increasingly broad band of large-scale wavelengths is not captured by the unconditional prior. This deficit is a consequence of the patch size choice, not a fundamental limitation of the framework: \ref{app:patch_size} shows that a larger patch size at
$Re_\tau = 1000$ recovers the large-scale structures accurately, at the cost of a wider patch support.

\begin{figure}[!ht]
    \centering
    \includegraphics[width=1.0\linewidth]{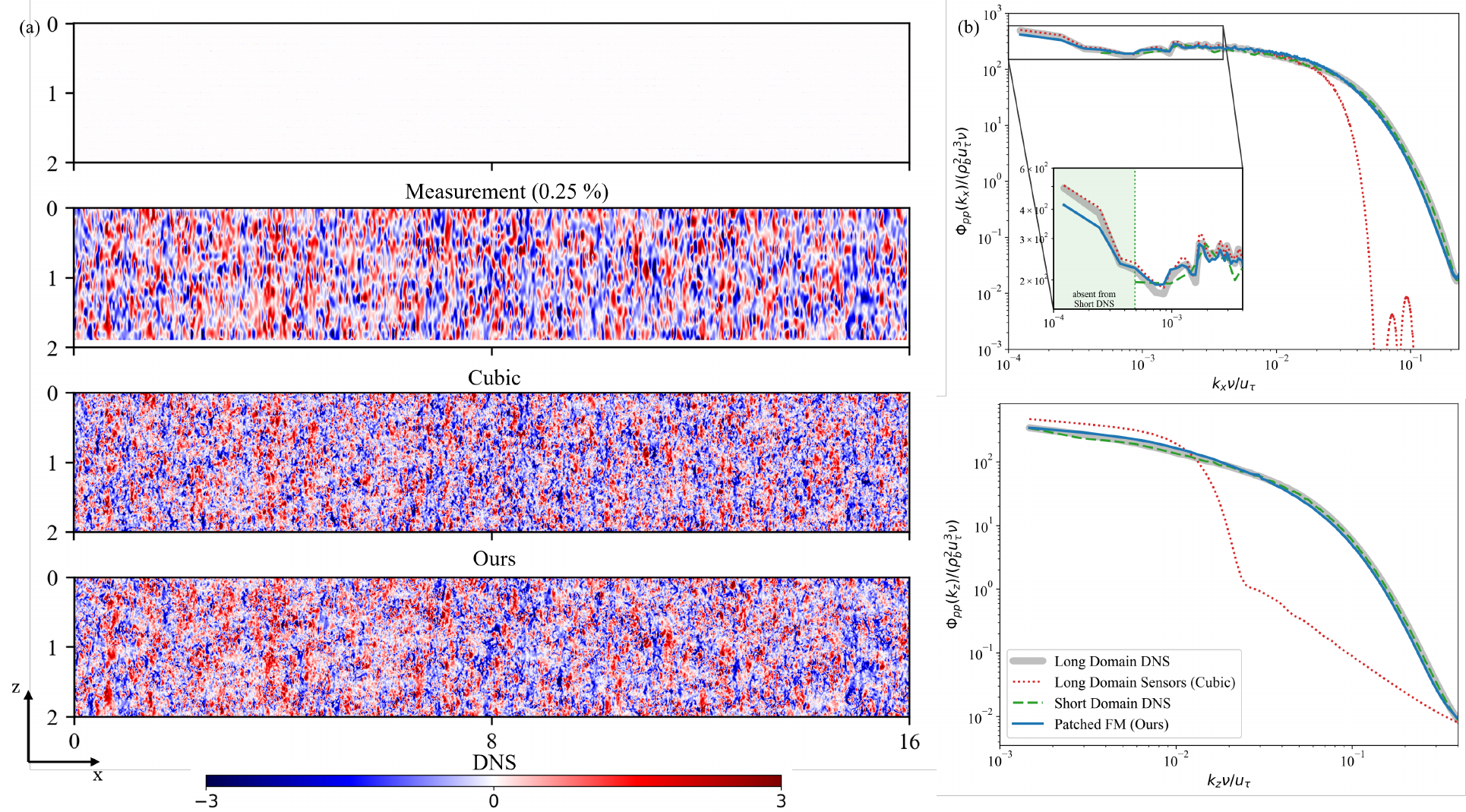}
    \caption{Conditional reconstruction at $Re_\tau=1000$ on the long domain ($L_x^L = 16\pi\delta$, Case~6): (a) Contours of the wall-pressure fluctuations from sparse sensor measurements ($0.25\%$ coverage), cubic interpolation, Patched FM reconstruction, and long-domain DNS; (b) streamwise and spanwise wavenumber spectra $\Phi_{pp}(k_x)$ and $\Phi_{pp}(k_z)$, comparing Patched FM, cubic interpolation, short-domain DNS, and long-domain DNS.}
    \label{fig:re1000_cond}
\end{figure}

The conditional reconstruction at $Re_\tau = 1000$ (Figure~\ref{fig:re1000_cond}) constitutes the most demanding test in this study: sensor coverage drops to $0.25\%$ of the long-domain grid. Even at this extreme sparsity, the sparse-sensor constraint restores the low-wavenumber streamwise content that the unconditional prior under-represents, and the Patched FM reconstruction recovers the spatial organization and spectral content of the long-domain DNS across the full resolved range, including features entirely absent from the short-domain DNS. The streamwise spectrum matches the long-domain DNS reference
closely, and the spanwise spectrum aligns with DNS throughout. Cubic interpolation deviates dramatically from DNS at all but the lowest wavenumbers, underscoring the inadequacy of conventional interpolation for reconstructing turbulent wall-pressure fields from sparse measurements at elevated Reynolds numbers.
Across all three Reynolds numbers, the results confirm that Patched FM, when trained on inner-scaled coordinates and extended to higher $Re_\tau$ via hierarchical transfer learning, provides a robust, data-efficient, and scalable approach for both unconditional generation and conditional reconstruction of wall-pressure fluctuations on domains significantly larger than the training configuration.

\section{Discussion}
\label{sec:discussion}

The results in Section~\ref{sec:results} established that Patched FM accurately generates wall-pressure fields on the short training domain and reconstructs fields with full-spectrum fidelity on a domain four times larger across three Reynolds numbers. In this section, we investigate the design choices that enable this capability through targeted comparisons and ablations, and probe the limits of the framework.

\subsection{Necessity of the patch-based decomposition}
\label{sec:baseline}

A natural question is whether the patch-based decomposition is necessary, or whether a standard flow matching model trained on the full spatial domain would suffice. To address this, we train a baseline FM model~\cite{lipman2022flow,parikh2026conditional} on full-resolution $256\times256$ snapshots from the short-domain DNS at $Re_\tau = 180$ (Case~1), where the basline processes the entire $256\times256$ field as the input. Both models are trained for an identical number of iterations, each seeing $1{,}200\text{k}$ total training snapshots. For a fair comparison, the baseline has a comparable parameter count (${\sim}31\,\text{M}$) to Patched FM (${\sim}29\,\text{M}$).

\begin{figure}[t!]
    \centering
    \includegraphics[width=0.8\linewidth]{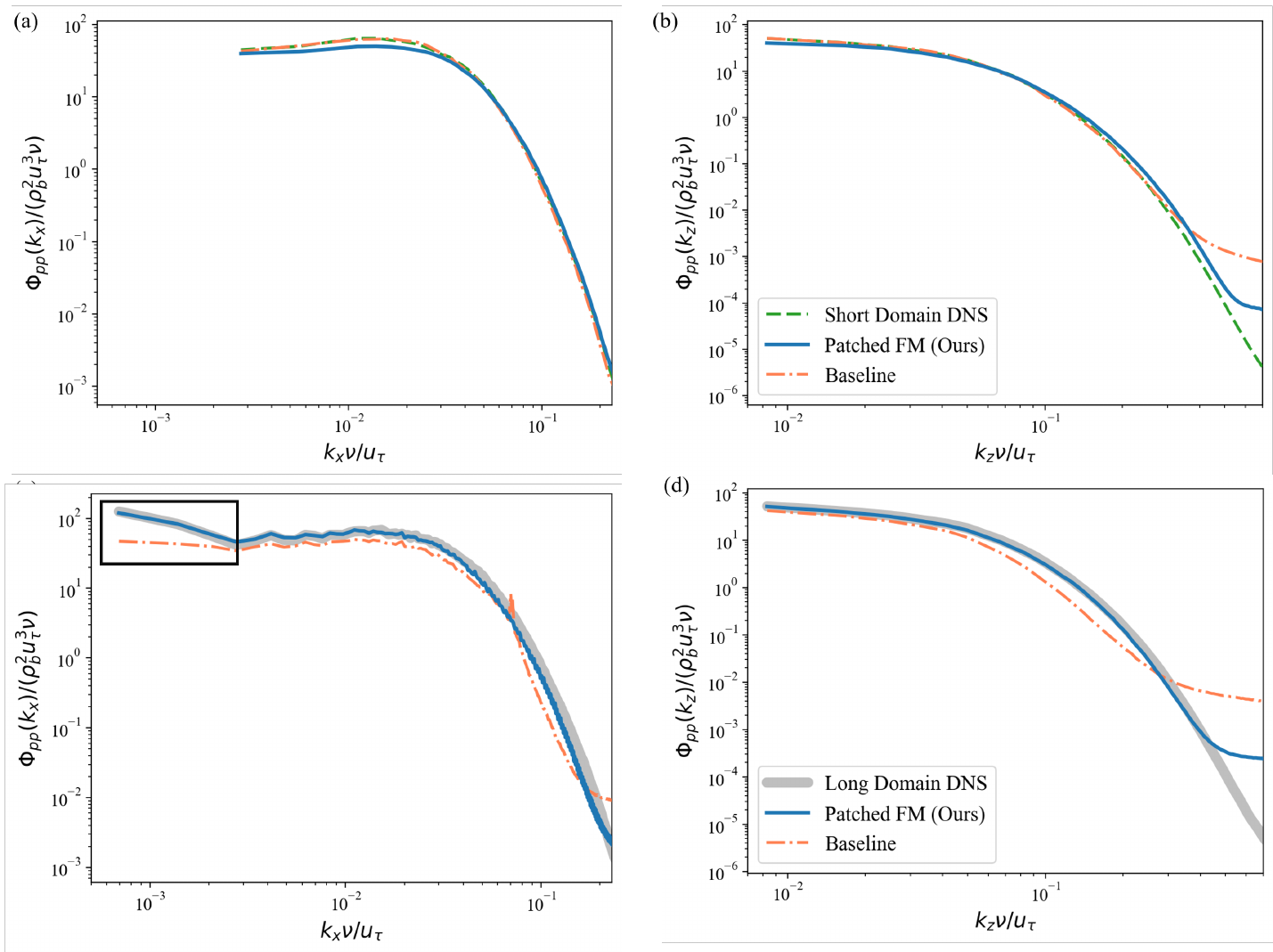}
    \caption{Comparison of Patched FM against a baseline flow matching model trained on full-resolution data at $Re_\tau=180$: (a,b) streamwise and spanwise wavenumber spectra on the training domain ($L_x^S=4\pi\delta$); (c,d) streamwise and spanwise wavenumber spectra on the testing domain ($L_x^L=16\pi\delta$). The baseline is conditioned on the same sparse sensor measurements as Figure~\ref{fig:re180_cond}(a).}
    \label{fig:comparison_baseline}
\end{figure}
Figure~\ref{fig:comparison_baseline} compares the two models spectrally. On the training domain (Figures~\ref{fig:comparison_baseline}(a,b)), the streamwise spectra of both models agree closely with the short-domain DNS reference. However, a notable difference already emerges in the spanwise direction: the baseline model exhibits a premature spectral roll-off at high spanwise wavenumbers, whereas Patched FM maintains agreement with DNS throughout the resolved range. This suggests that the patch-based decomposition provides a regularization effect that improves the representation of fine-scale spanwise structures even on the training domain. The critical distinction emerges on the testing domain ($L_x^L = 16\pi\delta$, Figures~\ref{fig:comparison_baseline}(c,d)), conditioned on the same sparse sensor measurements as Figure~\ref{fig:re180_cond}(a). In the streamwise direction (Figure~\ref{fig:comparison_baseline}(c)), the baseline under-estimates spectral density at the lowest wavenumbers (highlighted box), a direct consequence of the tiling strategy introducing spurious periodicity at multiples of the training domain length $4\pi\delta$. In the spanwise direction (Figure~\ref{fig:comparison_baseline}(d)), the baseline substantially over-estimates energy at intermediate and high wavenumbers. Patched FM, by contrast, matches the long-domain DNS reference accurately in both directions across the full resolved range. This comparison confirms that the patch-additive decomposition is essential for domain-size generalization: by decoupling the generative representation from the training domain, Patched FM avoids the tiling artifacts that corrupt the baseline's spectral content on the extended domain.

\subsection{Role of inner scaling in enabling Reynolds-number transfer}
\label{sec:inner-outer-scaled}

\begin{figure}[!ht]
    \centering
    \includegraphics[width=1.0\linewidth]{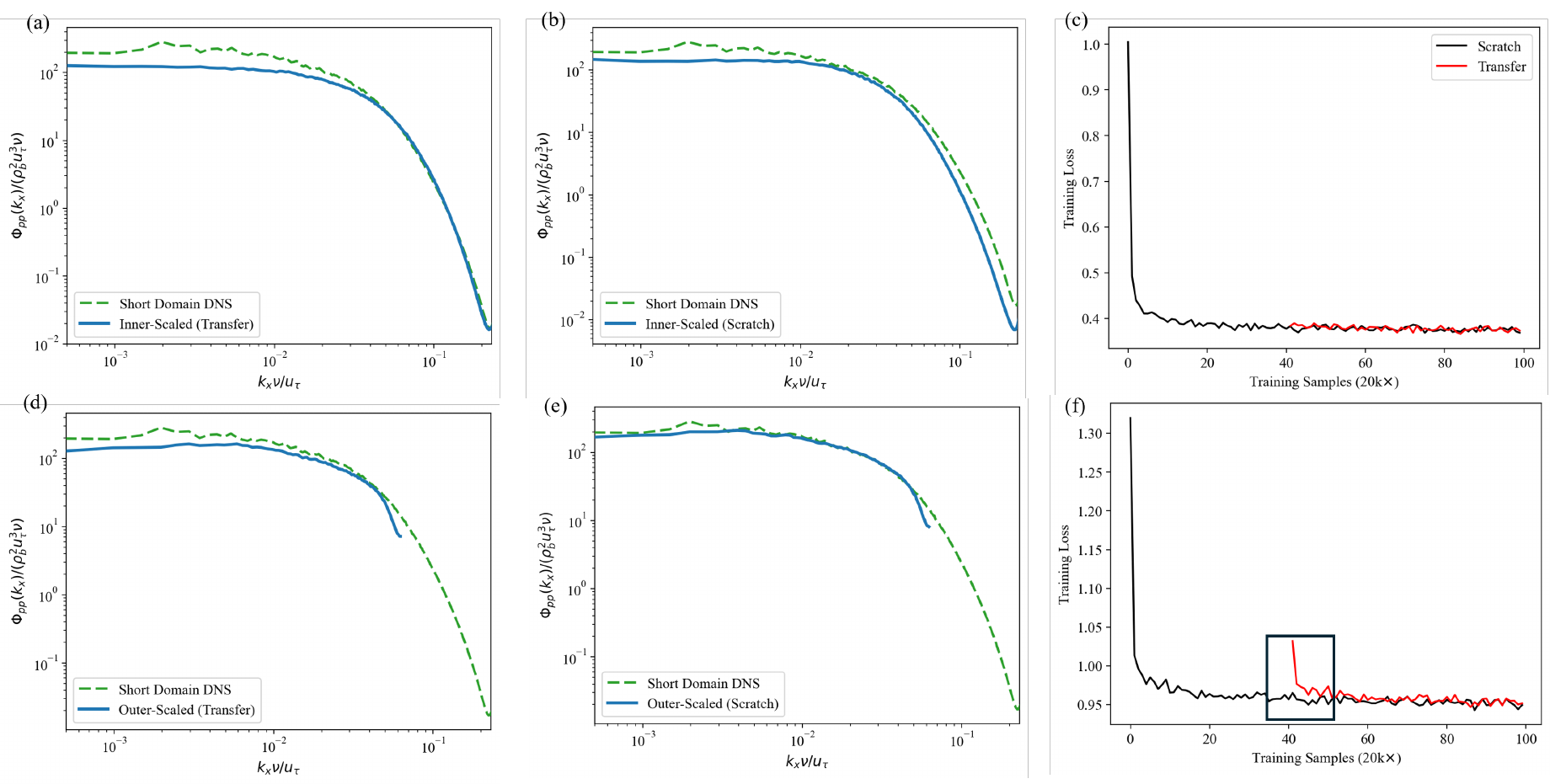}
    \caption{Comparison of inner-scaled versus outer-scaled Patched FM
    models at $Re_\tau=1000$: (a,d) transfer-learned models
    ($180\to1000$); (b,e) models trained from scratch; (c,f) training
    loss curves. Top row: inner-scaled coordinates; bottom row:
    outer-scaled coordinates.}
    \label{fig:comparison_inner_outer}
\end{figure}

The choice to scale the wall-pressure data in inner (viscous) units is motivated by the well-established universality of small-scale wall-pressure statistics when scaled by $u_\tau$ and
$\nu$~\cite{farabee1991spectral,klewicki2008statistical}. An alternative is to use outer units, which preserves the large-scale energy but causes the high-wavenumber content to vary with Reynolds number. The differences between these two scaling choices are illustrated in physical and spectral space in Figure~\ref{fig:motivation}; here we compare their effect on the Patched FM model at $Re_\tau = 1000$, examining both scratch-trained and transfer-learned ($180\to1000$) variants.
Before analyzing the results, a brief interpretive note on the training loss curves is necessary. The FM objective regresses the network onto a stochastic conditional velocity target defined by a given pairing of a training sample and Gaussian noise~\cite{lipman2022flow,tong2023improving}. Because the same intermediate state can be produced by many training samples, this target has an irreducible variance, and the loss therefore possesses a non-zero, \emph{a priori} unknown optimal value; it reflects only the
relative quality of the fit and is not a direct measure of absolute generative performance~\cite{xu2025diagnosing}. Two consequences follow. First, the global minimum of the empirical loss corresponds to the empirical velocity field, which collapses the reverse process onto the training samples; a model that generalizes instead converges to the marginal velocity field and its loss plateaus strictly above this empirical minimum~\cite{bonnaire2026diffusion}. Consequently, a loss that continues to decrease below the plateau is a signature of progression toward the memorizing solution rather than generation~\cite{buchanan2026edge}. Second, the denoising ability of the model can continue to improve even after the training loss has stagnated, so sample quality must be assessed from the generated fields rather than from the loss value alone~\cite{bonnaire2026diffusion,xu2025diagnosing}. This discussion is important for correctly interpreting the loss curves shown in Figures~\ref{fig:comparison_inner_outer} and~\ref{fig:limited_data}.

Figure~\ref{fig:comparison_inner_outer} presents the results. In the inner-scaled formulation (top row), the transfer-learned model (Figure~\ref{fig:comparison_inner_outer}(a)) reproduces the DNS spectrum accurately across moderate and high wavenumbers, and the training loss converges rapidly (Figure~\ref{fig:comparison_inner_outer}(c)), benefiting from the weights pre-trained at lower Reynolds numbers. By contrast, the
scratch-trained inner-scaled model (Figure~\ref{fig:comparison_inner_outer}(b)) does not reach the same level of spectral accuracy as its transfer-learned counterpart even after more training iterations, indicating that transfer learning successfully exploits the pre-trained low-Reynolds-number features to learn the high-Reynolds-number behavior.

In contrast, the outer-scaled formulation (bottom row) exhibits notable deficiencies in the intermediate-to-high wavenumber range. The transfer-learned outer-scaled model (Figure~\ref{fig:comparison_inner_outer}(d)) under-predicts energy at intermediate wavenumbers and exhibits a premature spectral roll-off at high wavenumbers, producing spectra that deviate substantially from DNS in precisely the range where inner-scaled statistics are expected to be universal. The scratch-trained outer-scaled model
(Figure~\ref{fig:comparison_inner_outer}(e)) shows similar deficiencies, indicating that the outer-scaled formulation struggles to represent the high-wavenumber content regardless of initialization. This behavior is expected: in outer-scaled coordinates, the high-wavenumber content varies with Reynolds number, so neither transfer-learned nor scratch-trained outer-scaled weights can reliably capture the viscous-scale statistics at $Re_\tau = 1000$. This interpretation is corroborated by the training loss curves
(Figure~\ref{fig:comparison_inner_outer}(f)). The transfer-learned outer-scaled model exhibits a pronounced loss spike at approximately $40\text{k}$ training samples before reconverging, whereas the scratch-trained model decreases monotonically. This spike is a direct signature of representation incompatibility: the outer-scaled weights pre-trained at $Re_\tau=180$ encode low-wavenumber features that are actively misleading at $Re_\tau=1000$, so the network must first unlearn these features, at the cost of a temporary loss increase, before it can adapt to the new multi-scale behavior. The net result is that the transfer-learned outer-scaled model effectively discards its pre-trained initialization and relearns
from scratch, providing no benefit over random initialization. By contrast, the inner-scaled transfer loss in panel (c) decreases monotonically from the start, confirming that the pre-trained inner-scaled features are directly compatible with the target Reynolds number and require only incremental adaptation rather than wholesale overwriting.

\subsection{Performance under limited data}
\label{sec:limited_data}

Having established that inner scaling enables efficient Reynolds-number transfer, we now probe how far the data requirement can be reduced. High-fidelity DNS data at elevated Reynolds numbers are scarce and expensive to generate; it is therefore important to assess whether the framework retains generation quality under severe data constraints. This question is especially relevant given the known tendency of generative models to overfit when trained on small datasets~\cite{xu2025diagnosing,bonnaire2026diffusion,buchanan2026edge}.

\begin{figure}[!ht]
    \centering
    \includegraphics[width=1.0\linewidth]{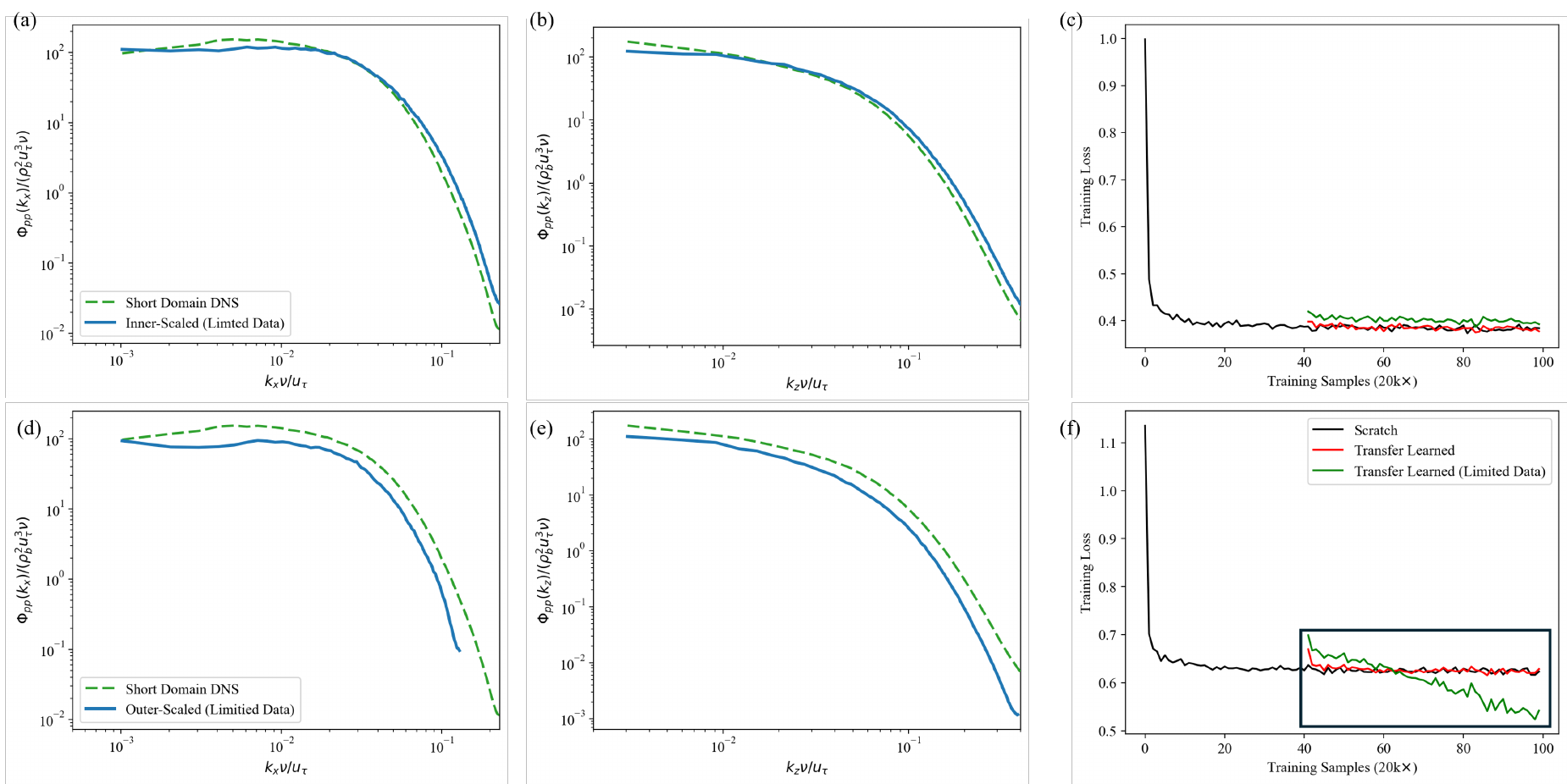}
    \caption{Generation performance under severely limited training data ($10$ snapshots) at $Re_\tau=500$: (a,d) streamwise and (b,e) spanwise power spectral densities for inner-scaled (top row) and outer-scaled (bottom row) transfer-learned models; (c,f) training loss curves comparing scratch training, standard transfer learning, and transfer learning with limited data. A loss descending below the generalization plateau (marginal-velocity solution) indicates progression toward memorization rather than improved generation~\cite{bonnaire2026diffusion}.}
    \label{fig:limited_data}
\end{figure}
Figure~\ref{fig:limited_data} evaluates the model pre-trained at $Re_\tau=180$ and fine-tuned at $Re_\tau=500$ using only $10$ short-domain snapshots---a $50\times$ reduction relative to the $500$ snapshots used in Section~\ref{sec:hie-transfer-learning} and a $98\%$ reduction relative to the base training set. In the inner-scaled formulation (top row), the limited-data transfer-learned model (Figures~\ref{fig:limited_data}(a),(b)) continues to produce spectra in close agreement with DNS in both directions. The training loss (Figure~\ref{fig:limited_data}(c)) plateaus at the marginal-velocity solution and does not descend toward the empirical minimum, confirming that the pre-trained inner-scaled weights supply sufficient physical structure for the network to generalize from extremely few target-$Re_\tau$ examples.

The outer-scaled formulation (bottom row) shows a sharply contrasting
behavior. The outer-scaled limited-data model (Figures~\ref{fig:limited_data}(d),(e)) exhibits pronounced spectral deviations from DNS, and its training loss (Figure~\ref{fig:limited_data}(f), highlighted region) descends below
the marginal-velocity plateau reached by both the scratch-trained and
fully-trained transfer-learned models, which is the signature of memorization
of the $10$-snapshot training set. This failure follows directly from the analysis in Section~\ref{sec:inner-outer-scaled}: the outer-scaled pre-trained features are actively misleading at the new Reynolds number, so the model must overwrite them from insufficient data.

Taken together, Sections~\ref{sec:inner-outer-scaled} and~\ref{sec:limited_data} establish a consistent picture: inner scaling makes the pre-trained representation physically compatible with the target Reynolds number, which is both necessary for successful transfer and sufficient to reduce the data requirement to a level that is practically accessible (as few as $10$ snapshots) without sacrificing generation quality.

\subsection{Zero-shot generalization across Reynolds numbers}

The ultimate test of the transferability of the learned patch prior is zero-shot inference: applying a model trained at one Reynolds number to reconstruct fields at an entirely different, unseen Reynolds number without any fine-tuning. This tests whether the inner-scaled representation has learned genuinely universal wall-pressure statistics, or merely Reynolds-number-specific features that happen to generalize within a narrow range.

\begin{figure}[!ht]
    \centering
    \includegraphics[width=1.0\linewidth]{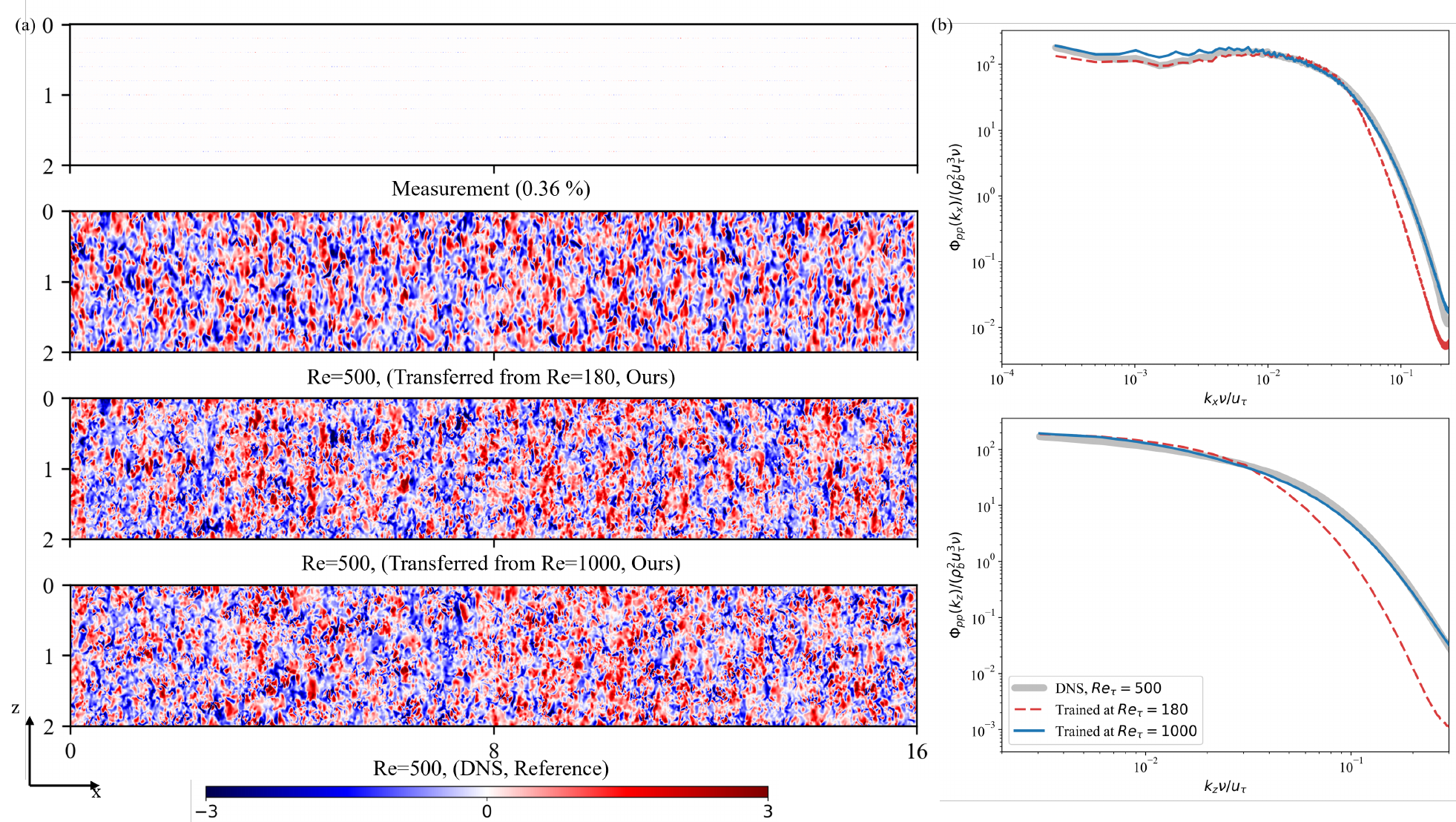}
    \caption{Zero-shot conditional reconstruction at $Re_\tau=500$
    without any fine-tuning at the target Reynolds number:
    (a) Contours of the wall-pressure fluctuations from sparse sensor measurements
    ($0.36\%$ coverage), zero-shot reconstructions using models
    trained at $Re_\tau=180$ and $Re_\tau=1000$, and long-domain DNS
    ground truth ($L_x^L=16\pi\delta$, Case~5); (b) streamwise (top)
    and spanwise (bottom) wavenumber spectra $\Phi_{pp}(k_x)$ and $\Phi_{pp}(k_z)$ comparing the two
    zero-shot reconstructions against long-domain DNS at
    $Re_\tau=500$.}
    \label{fig:zero_shot}
\end{figure}

Figure~\ref{fig:zero_shot} presents zero-shot conditional reconstructions of the $Re_\tau=500$ long-domain field using models trained exclusively at $Re_\tau=180$ and $Re_\tau=1000$, conditioned on $0.36\%$ sensor coverage from the long-domain DNS.

The $Re_\tau=1000$ model produces reconstructions that closely match the DNS ground truth both visually and spectrally. The generated fields reproduce the spatial organization, multi-scale structures, and intermittency of the $Re_\tau=500$ DNS reference (Figure~\ref{fig:zero_shot}(a)), and the streamwise and spanwise spectra align with DNS across the full resolved range (Figure~\ref{fig:zero_shot}(b)). This result is consistent with the inner-scaling hypothesis: in inner-scaled coordinates, the $Re_\tau=500$ and $1000$ cases share nearly identical high-wavenumber statistics, so the patch prior trained at $Re_\tau=1000$ is directly applicable at $Re_\tau=500$ without modification. The sparse sensor constraint anchors the large-scale low-wavenumber content, and the pre-trained patch prior supplies the small-scale structures with no additional adaptation required.

The $Re_\tau=180$ model, by contrast, produces fields that visually recover the large-scale organization but carry insufficient small-scale energy relative to the DNS reference
(Figure~\ref{fig:zero_shot}(a)). Spectrally, both the streamwise and spanwise spectra roll off earlier than DNS at intermediate-to-high wavenumbers (Figure~\ref{fig:zero_shot}(b)). This is expected: while the high-wavenumber decay is approximately universal in inner scaling, the $Re_\tau=180$ field carries less energy in the intermediate wavenumber range than the $Re_\tau=500$ target, so a patch prior built from $Re_\tau=180$ data systematically biases the unconditional content of the reconstruction toward lower energy levels. The sparse-sensor likelihood compensates for the low-wavenumber content, which is why the large-scale spatial organization is still captured; nonetheless, the residual intermediate-band energy deficit persists because it falls outside the wavenumber range that the sensor constraint directly controls. Even so, the overall spectral shape is
reproduced reasonably well: the model captures the correct slope and decay trend, with only a downward bias in amplitude. This partial success explains why hierarchical transfer learning from $Re_\tau=180$ is an effective initialization strategy: the pre-trained prior already encodes the correct spectral structure, and fine-tuning on a small number of target-$Re_\tau$ snapshots needs only to correct the energy level rather than relearn the spectral shape from scratch.


\subsection{Training and inference computational costs}
\label{sec:train_and_inf_costs}

The patch-based formulation provides computational advantages at both training and inference time that compound with the hierarchical transfer-learning strategy.

\paragraph{Training cost}
Because $v_\theta$ is trained exclusively on fixed-size $P\times P$ patches extracted from the short-domain DNS, the training cost is entirely independent of the target domain size; the same trained model is deployed on domains of arbitrary extent at inference time
without retraining. A conventional generative model trained on full-resolution fields, by contrast, would require retraining whenever the target domain changes, with a training cost scaling with the number of grid points. For the $Re_\tau=1000$ long-domain case (Case~6), this would entail processing a $4800\times800$ field, a $4\times$ increase in input dimension relative to the short-domain counterpart. The patch-based formulation entirely avoids this scaling by decoupling the generative representation from the domain extent.

\begin{figure}[!ht]
    \centering
    \includegraphics[width=1.00\linewidth]{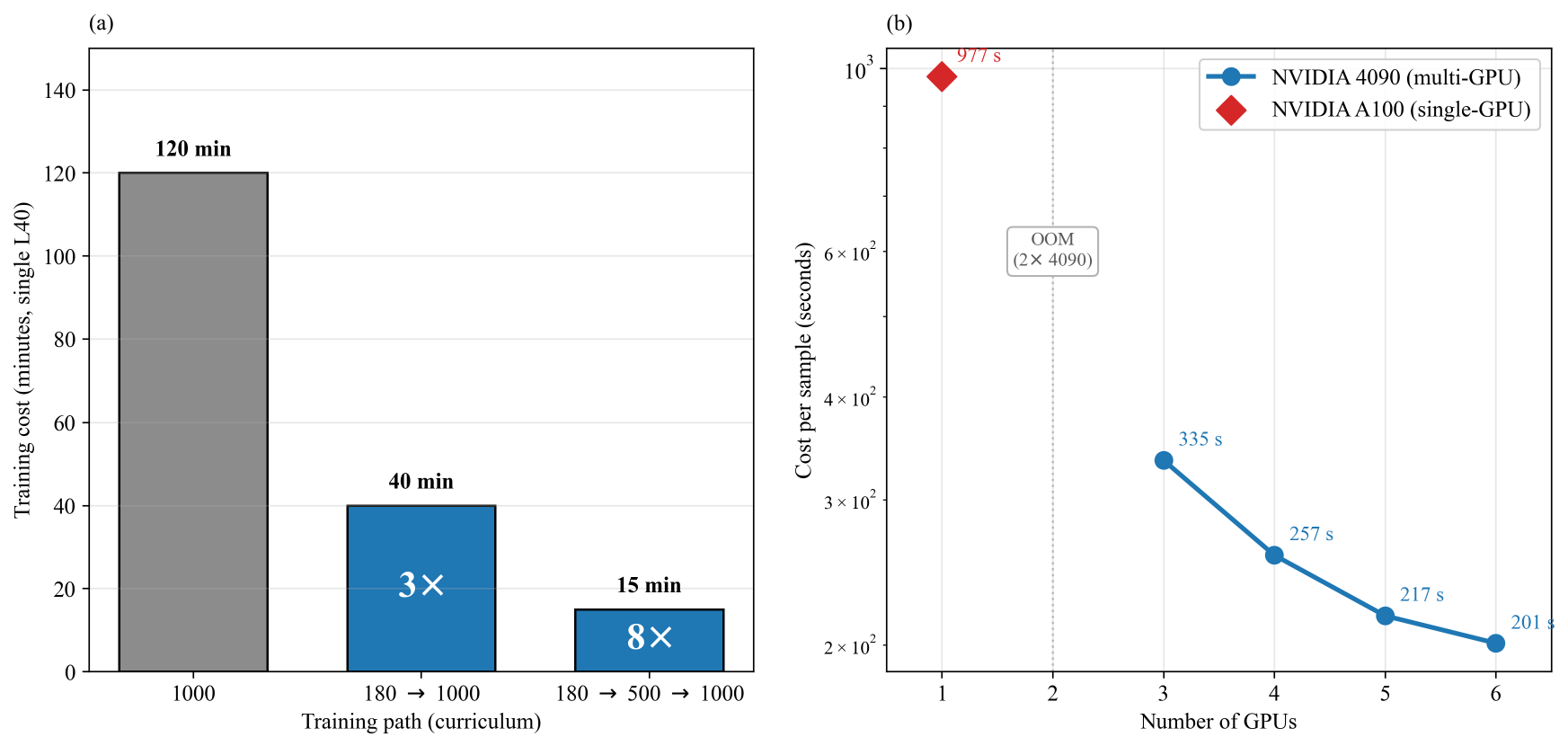}
    \caption{Computational cost of the Patched FM framework.
    (a) Cumulative training time (minutes, single NVIDIA L40 GPU)
    for three training strategies at $Re_\tau=1000$: scratch training
    ($120$\,min), direct transfer learning ($180\to1000$, $40$\,min),
    and the proposed hierarchical transfer learning
    ($180\to500\to1000$, $15$\,min), corresponding to $8\times$ and
    $3\times$ reductions relative to scratch training and direct
    transfer, respectively.
    (b) Per-sample wall-clock inference time for conditional
    generation at $Re_\tau=1000$ ($4800\times800$ grid) as a function
    of GPU count; single-GPU inference uses an NVIDIA A100 80\,GB,
    multi-GPU inference uses NVIDIA 4090 24\,GB cards, and the
    $2\times$4090 configuration runs out of memory (OOM).}
    \label{fig:train_and_inf_costs}
\end{figure}

The hierarchical transfer-learning strategy further reduces the training cost at elevated Reynolds numbers. Figure~\ref{fig:train_and_inf_costs}(a) shows that hierarchical fine-tuning ($180\to500\to1000$, $15$\,min) is $8\times$ cheaper than training from scratch ($120$\,min) and $3\times$ cheaper than direct transfer ($180\to1000$, $40$\,min) on a single NVIDIA L40 GPU. The cost advantage of the hierarchical schedule over direct
transfer is a direct consequence of the inner-scaling property established in Section~\ref{sec:inner-outer-scaled}: by fine-tuning through intermediate Reynolds numbers, the model adapts incrementally to the new scales without overwriting pre-trained
features that remain valid at the target $Re_\tau$. These training cost advantages compound with the data efficiency demonstrated in Section~\ref{sec:limited_data}, making the overall training pipeline practical even when high-$Re_\tau$ DNS data are scarce.

\paragraph{Inference cost}
The inference workload is highly parallelizable. During conditional generation, all patches within each autoregressive section are propagated through the flow-matching ODE in a single batched forward pass and can be distributed across parallel hardware; the long domain ($L_x^L = 16\pi\delta$) is reconstructed as four consecutive sections of $4\pi\delta$ each, with inter-section coherence maintained by the spatial consistency constraint (Section~\ref{sec:inference}).
Figure~\ref{fig:train_and_inf_costs}(b) reports the per-sample wall-clock cost for the most computationally demanding case in this study, the $Re_\tau = 1000$ conditional generation on the $4800\times800$ long-domain grid (Case~6), as a function of GPU
count. A single NVIDIA A100 80\,GB GPU requires $977$\,s per sample; a distributed multi-GPU implementation on NVIDIA 4090 24\,GB cards reduces this to $201$\,s on six GPUs, a nearly five-fold speedup, with diminishing returns beyond four GPUs. The $2\times4090$
configuration runs out of memory, illustrating the VRAM bottleneck that the patch-distributed strategy relieves by allowing subsets of patches to reside on different devices. This parallel scalability stands in direct contrast to monolithic generative models, whose inference cost and memory requirement scale with the total number
of grid points and cannot be reduced by patch-level distribution.

\section{Conclusion}
\label{sec:conclusion}

We propose Patched FM, a generative-prior framework that exploits the inner-scaling self-similarity of small-scale wall-pressure fluctuations in wall-bounded turbulence to generalize across both longer spatial domains and higher friction Reynolds numbers $Re_\tau$. To demonstrate its advantages, we apply it to the recovery of large-scale structures absent from the training set, using a test-time, training-free, spatially autoregressive conditional generation scheme that assimilates information from the sensors and from previously generated domains. This approach blends sensor measurements with multiscale physics while avoiding the prohibitive computational cost of long-domain simulations. We further show that the Patched FM prior can be fine-tuned to higher $Re_\tau$ in a data-efficient, hierarchical manner, establishing the scalability of the method. We then provide a detailed analysis of the framework, isolating the role of its central assumptions: the suitability of a patchwise generative prior for this physical problem, and the necessity of inner-scaled coordinates for generalizing across $Re_\tau$, which we establish by contrast with models trained on outer-scaled coordinates. Finally, we highlight two further properties of Patched FM: highly data-efficient training and zero-shot generalization from higher to lower $Re_\tau$.

\section*{Acknowledgements}
The authors are grateful to Meng Wang for sharing the DNS dataset and for insightful discussions. This work was supported by the U.S. Office of Naval Research under award number N00014-23-1-2071.

\bibliographystyle{elsarticle-num}
\bibliography{ref}

\appendix

\section{Supplementary Studies}

\subsection{Effect of number of sensors on generation performance}
\label{app:sensor_ablation}

The conditional generation framework relies on sparse sensor observations to guide the reconstruction of wall-pressure fields on the extended domain. The number and density of these sensors directly influences the quality of the reconstruction, particularly the recovery of low-wavenumber content that requires long-range spatial information.

\begin{figure}[!ht]
    \centering
    \includegraphics[width=1.0\linewidth]{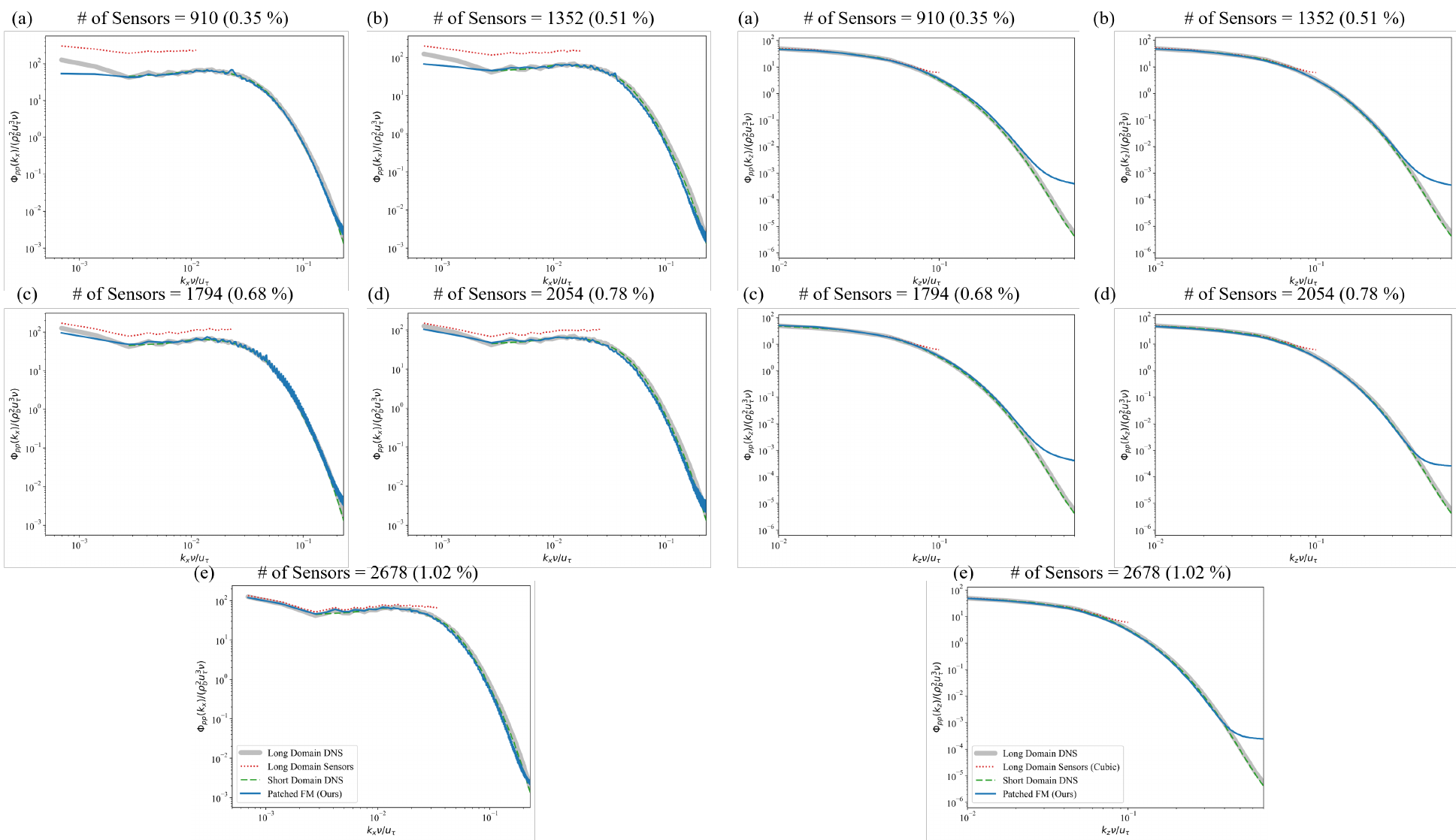}
    \caption{Ablation study on the effect of sensor density on conditional generation quality at $Re_\tau=180$: streamwise (left) and spanwise (right) wavenumber spectra of the wall-pressure fluctuations, $\Phi_{pp}(k_x)$ and $\Phi_{pp}(k_z)$, for varying numbers of sensors. Results are compared against short- and long-domain DNS references.}
    \label{fig:sensor_ablation}
\end{figure}

Figure~\ref{fig:sensor_ablation} presents an ablation study at $Re_\tau = 180$, where the sensor count is systematically varied. At extremely low sensor densities, the reconstructed streamwise spectrum exhibits deviations from the long-domain DNS at low wavenumbers, as the guidance signal provides insufficient constraint on the inter-patch coherence. As the number of sensors increases, the streamwise spectrum progressively converges toward the DNS reference, with the low-wavenumber rise being captured once the sensor spacing becomes comparable to or smaller than the dominant wavelength of the large-scale structures. In the spanwise direction, the spectrum is less sensitive to the sensor count, since the spanwise patch dimension already resolves the relevant scales.

These results suggest that a moderate sensor density---on the order of 1\% of the total grid points---is sufficient for accurate reconstruction across all wavenumbers. Beyond this threshold, additional sensors yield diminishing returns. This finding is practically important, as it implies that the Patched FM framework can produce high-fidelity reconstructions from very sparse measurement arrays, as might be available in experimental settings.

\subsection{Effect of patch size}
\label{app:patch_size}

Although the Patched FM formulation (Section~\ref{sec:patched_fm}) guarantees a globally coherent field while training the generative prior only on patches, an aggressively small patch size (relative to the global field) can degrade the predictions and discard flow structures larger than the training patch. This is expected because the patch size $P$ imposes a hard ceiling on the wavelengths that the unconditional patch prior can represent: any feature whose support exceeds $P$ grid points cannot be expressed by the patch density $p_{i,j,B}$ in Equation~\eqref{eq:patch_product}. As noted in Section~\ref{sec:setup}, we use the randomized, mixed-patch-size training strategy $\mathbb{P} = \{56, 32, 16\}$ throughout this work, because our objective is to focus the patch prior on the high-wavenumber content of the wall-pressure field, where the inner-scaling self-similarity argument of Section~\ref{sec:problem} is strongest (see Figure~\ref{fig:motivation}). As shown in Section~\ref{sec:hie-transfer-learning}, at the higher Reynolds numbers $Re_\tau = \{500, 1000\}$ the low-wavenumber features that exceed the patch extent are not captured with sufficient accuracy by the unconditional prior; they are instead handled explicitly by the conditional sensor likelihood at inference time (Section~\ref{sec:inference}).

\begin{figure}[h]
    \centering
    \includegraphics[width=1.0\linewidth]{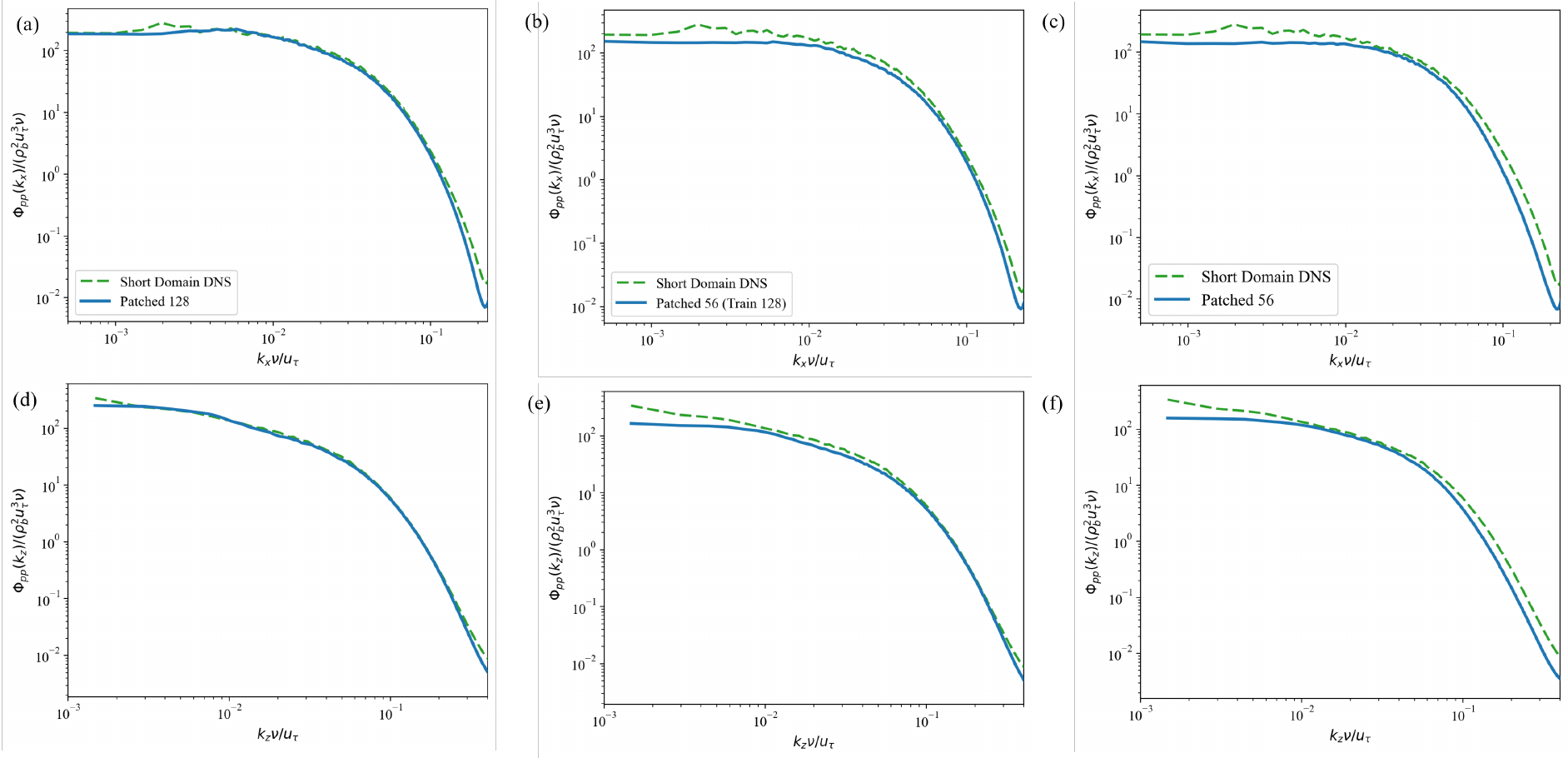}
    \caption{Sensitivity study on patch size at $Re_\tau = 1000$: streamwise (top) and spanwise (bottom) wavenumber spectra of the wall-pressure fluctuations, $\Phi_{pp}(k_x)$ and $\Phi_{pp}(k_z)$, for (a,d) a model trained with patch-size set $\mathbb{P}_{\mathrm{b}} = \{128, 56, 32\}$ and evaluated at patch size $128$; (b,e) a model trained with patch-size set $\mathbb{P}_{\mathrm{b}} = \{128, 56, 32\}$ but evaluated at patch size $56$; and (c,f) a model trained with patch-size set $\mathbb{P}_{\mathrm{s}} = \{56, 32, 16\}$ and evaluated at patch size $56$.}
    \label{fig:patch_size}
\end{figure}

To illustrate the performance trade-off introduced by the patch size, particularly for the larger-$Re_\tau$ cases, we train two Patched FM priors using the patch-size sets $\mathbb{P}_{\mathrm{b}} = \{128, 56, 32\}$ and $\mathbb{P}_{\mathrm{s}} = \{56, 32, 16\}$, under a fixed budget of $1200$k total training images. The model architecture is identical to the settings in \ref{app:unet}. Figure~\ref{fig:patch_size} reports three configurations: (a,d) the $\mathbb{P}_{\mathrm{b}}$ prior evaluated at patch size $128$; (b,e) the $\mathbb{P}_{\mathrm{b}}$ prior evaluated at patch size $56$; and (c,f) the $\mathbb{P}_{\mathrm{s}}$ prior evaluated at patch size $56$. We refer to these as configurations (A), (B), and (C), respectively.

In the streamwise direction, comparing configuration (A) (Figure~\ref{fig:patch_size}(a)) with configuration (C) (Figure~\ref{fig:patch_size}(c)) shows that (C) produces a spectrum with a pronounced low-wavenumber discrepancy relative to the DNS, indicating that the streamwise energy content of interest at $Re_\tau = 1000$ is adequately captured at a patch size of $128$. The spanwise spectra (Figure~\ref{fig:patch_size}(d) and (f)) exhibit the same trend.

Although larger patches yield more accurate results for a fixed number of training iterations, they are more memory- and compute-intensive owing to their larger dimensions. We therefore introduce configuration (B), which is trained on the larger patch size but evaluated on the smaller one, alleviating the memory and computational bottlenecks at inference. Figure~\ref{fig:patch_size}(b) and (e) show that this strategy outperforms configuration (C) and is only partially degraded relative to configuration (A).

\subsection{UNet architecture and training hyperparameters}
\label{app:unet}

The velocity network $v_\theta$ is a UNet adapted from the architecture used in \cite{parikh2026conditional}. The network is multi-resolution, with residual blocks at each level, attention blocks at the two coarsest resolutions ($P/2$ and $P/4$, including the bottleneck), and a sinusoidal time embedding for the flow-matching time variable $s$. The input to the network is a three-channel patch tensor consisting of the wall-pressure value $\phi_s$ together with two spatial-coordinate channels $X$ and $Z$ (cf.\ Figure~\ref{fig:schematic}); a single shared network is used across all patches and across all Reynolds numbers considered.

Specific architectural and training hyperparameters used in this work are summarized in Table~\ref{tab:unet}.

\begin{table}[!ht]
    \centering
    \caption{UNet architecture and training hyperparameters used in this work.}
    \label{tab:unet}
    \begin{tabular}{ll}
        \hline
        Hyperparameter & Value \\
        \hline
        Trainable parameters & $\sim29$M \\
        Input/output channels & $3$ \\
        Input/output patch size & $P \in \{16, 32, 56\}$ \\
        Base channels & $128$ \\
        Base channel multiplier & $(1, 2, 2)$ \\
        Residual blocks per level & $2$ \\
        Attention resolutions & $(P/2,\, P/4)$ \\
        Attention head channels & $64$ \\
        Time embedding dimension & $512$ \\
        Optimizer & Adam \\
        Learning rate & $10^{-4}$ \\
        Batch size & $256$ \\
        ODE solver & Euler \\
        Number of ODE steps & $100$ \\
        \hline
    \end{tabular}
\end{table}

\end{document}